\documentclass[dvipsnames]{article}
\usepackage{unipat_report}

\usepackage[utf8]{inputenc}
\usepackage[T1]{fontenc}
\usepackage{booktabs}
\usepackage{amsfonts}
\usepackage{amsmath}
\usepackage{nicefrac}
\usepackage{colortbl}
\usepackage{graphicx}
\usepackage{float}
\usepackage{multirow}
\usepackage{makecell}
\usepackage[inline]{enumitem}
\usepackage[skins,breakable]{tcolorbox}
\usepackage{pifont}
\usepackage{subcaption}
\usepackage{placeins}
\newcommand{\cmark}{\ding{51}}
\newcommand{\xmark}{\ding{55}}

\tcbset{
  taskbox/.style={
    enhanced, breakable,
    colframe=blue!55!black,
    colback=blue!2!white,
    colbacktitle=blue!55!black,
    coltitle=white,
    fonttitle=\small\bfseries,
    fontupper=\small,
    boxrule=0.8pt,
    arc=3pt,
    top=5pt, bottom=5pt, left=7pt, right=7pt,
    before upper={\setlength{\parindent}{0pt}\setlength{\parskip}{0pt}},
  },
  phasebox/.style={
    enhanced,
    colback=white,
    colbacktitle=blue!48!black,
    coltitle=white,
    colframe=blue!35!black,
    fonttitle=\small\bfseries,
    boxrule=0.4pt,
    arc=2pt,
    top=2pt, bottom=3pt, left=5pt, right=5pt,
    toptitle=2pt, bottomtitle=2pt,
    before={\vspace{4pt}},
    before upper={\setlength{\parindent}{0pt}\setlength{\parskip}{2pt}},
  }
}

\newcommand{\MYBENCH}{\textsc{RoadmapBench}}

\title{\MYBENCH: Evaluating Long-Horizon Agentic Software Development Across Version Upgrades}

\author{
  \small{Xinbo~Xu$^{1,2}$, Ruihan~Yang$^{3}$, Haiyang~Shen$^{1,2}$, Wendong~Xu$^{1,4}$, Bofei~Gao$^{2}$, Ruoyu~Wu$^{1,2}$, Kean~Shi$^{1,2}$, Weichu~Xie$^{2}$, Xuanzhong~Chen$^{1,5}$, Ming~Wu$^{6}$, Jason~Zeng$^{6}$, Michael~Heinrich$^{6}$, Elvis~Zhang$^{7}$, Liang~Chen$^{1\dagger}$, Kuan~Li$^{1\dagger}$, Baobao~Chang$^{2\dagger}$}
  \\[0.8em]
  {\fontsize{10pt}{11pt}\selectfont
  $^1$UniPat AI \quad $^2$Peking University \quad $^3$Fudan University\\[2pt]
  $^4$The University of Hong Kong \enspace $^5$Tsinghua University \enspace $^6$0G Labs \enspace $^7$Pipeline Lab}\\
}

\begin{document}

\maketitle
\renewcommand{\thefootnote}{\fnsymbol{footnote}}
\footnotetext[2]{Corresponding authors: \texttt{liangchen@unipat.ai}, \texttt{kuanli@unipat.ai}, \texttt{chbb@pku.edu.cn}}
\renewcommand{\thefootnote}{\arabic{footnote}}

\begin{abstract}
Coding agents are increasingly deployed in real software development, where a single version iteration requires months of coordinated work across many files. However, most existing benchmarks focus predominantly on single-issue bug fixes from Python repositories, with coarse pass/fail evaluation outcomes, and thus fail to capture long-horizon, multi-target development at real engineering scale.
To address this gap, we present \MYBENCH, a benchmark of 115 long-horizon coding tasks grounded in real open-source version upgrades across 17 repositories and 5 programming languages.
Each task places the agent on a source-version code snapshot and provides a multi-target roadmap instruction requiring it to implement the functionality introduced in the target version, with a median modification of 3{,}700 lines across 51 files.
We conduct a systematic evaluation on thirteen frontier models and find that even the strongest, Claude-Opus-4.7, resolves only 39.1\% of tasks, while the weakest achieves merely 5.2\%, in stark contrast to existing bug-fix benchmarks, suggesting that long-horizon software development remains a largely unsolved problem.
\end{abstract}

\begin{figure}[H]
  \centering
  \includegraphics[width=0.91\linewidth]{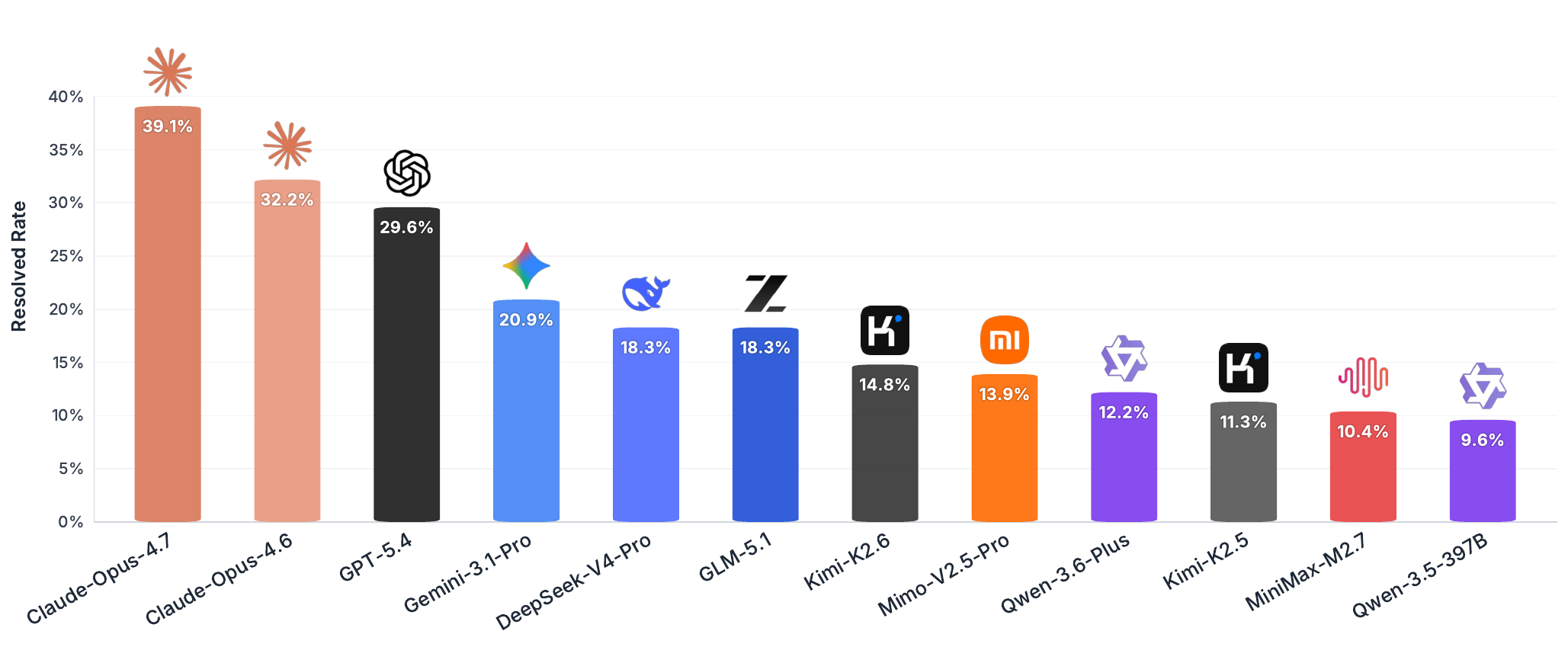}
  \vspace{-2mm}
  \caption{\textbf{RoadmapBench Leaderboard.} Resolved rate of top-performing models evaluated with OpenHands across 115 multi-target software evolution tasks spanning 5 languages and 17 repositories. Even the best-performing model resolves only 39.1\% of tasks.}
  \label{fig:leaderboard}
\end{figure}

\section{Introduction}

\begin{table}[t]
  \caption{Comparison with related coding benchmarks. Scope: task granularity. Subtask Score: target-level completion scoring. Solution: oracle patch size (LoC).}
  \vspace{6pt}
  \label{tab:comparison}
  \centering
  \small
  \setlength{\tabcolsep}{8.8pt}
  \begin{tabular}{lccccc}
    \toprule
    \textbf{Benchmark} & \textbf{\#Tasks} & \textbf{Lang.} & \textbf{Scope} & \textbf{Subtask Score} & \textbf{Solution} \\
    \midrule
    SWE-bench Verified~\cite{openai2024sweverified}  & 500   & Python & Commit          & \textcolor{red}{\xmark} & $\sim$33 LOC \\
    SWE-bench Pro~\cite{deng2025swe}                 & 1,865 & Multi  & Commit        & \textcolor{red}{\xmark} & $\sim$107 LOC \\
    FeatureBench~\cite{zhou2025featurebench}         & 200   & Python & Commit        & \textcolor{red}{\xmark} & $\sim$790 LOC \\
    TerminalBench~\cite{terminalbench2026}           & 89    & Multi  & Task  & \textcolor{red}{\xmark} & $\sim$280 LOC \\
    SWE-EVO~\cite{thai2025swe}                       & 48    & Python & Version        & \textcolor{red}{\xmark} & $\sim$611 LOC \\
    NL2Repo~\cite{ding2025nl2repo}                   & 104   & Python & Repo           & \textcolor{red}{\xmark} & $\sim$3,000 LOC \\
    \midrule
    \rowcolor{blue!6}
    \textbf{ROADMAPBENCH} & \textbf{115} & \textbf{Multi} & \textbf{Version} & \textcolor{green!60!black}{\cmark}\ \textbf{(avg.\ 5 targets)} & $\sim$\textbf{3,700 LOC} \\
    \bottomrule
  \end{tabular}
\end{table}

The rapid progress of large language models (LLMs)~\citep{anthropic2025claude46, openai2026gpt54, googledeepmind2025gemini3flash} has enabled a new generation of coding agents that can plan, edit, execute, and validate software in interactive development environments~\citep{yang2024swe,zhang2024codeagentenhancingcodegeneration,huang2025opencoder, wang2025swe}.
As these agents move beyond isolated code generation and bug fixing, the central challenge increasingly lies in sustained, multi-target software development.
Evaluation is therefore shifting from \textbf{short-horizon} defect repair to \textbf{long-horizon} feature implementation. This raises a critical question: \textit{how to evaluate an agent on multi-target, human-scale development work spanning weeks or months?}

Existing benchmarks have not kept pace with this shift (Table~\ref{tab:comparison}).
Most current benchmarks remain short-horizon: SWE-bench~\citep{jimenez2024swebench} and SWE-bench Pro~\citep{deng2025swe} evaluate isolated software engineering problems, with oracle solutions of $\sim$33 and $\sim$107 lines respectively, one to two orders of magnitude below the scale of real engineering work.
Long-horizon attempts remain scarce and collapse each task to a single binary outcome, overlooking the multi-target structure that real version upgrades naturally exhibit, where developers coordinate multiple substantial changes within a single release cycle.
Beyond scope and granularity, existing benchmarks remain concentrated in a limited set of heavily reused Python repositories~\citep{liu2023repobench, du2023classeval}, compounding contamination risk as popular codebases become more likely to appear in pre-training corpora.

To tackle these challenges, we propose \MYBENCH~, a benchmark of 115 long-horizon coding tasks grounded in real open-source version upgrades.
Each task starts from a repository snapshot pinned to an earlier release and requires the agent to implement the behaviors introduced in the next release, with a median oracle modification of approximately 3{,}700 lines across multiple files and modules.
We convert each upgrade into a multi-target roadmap with a median of 5 subtasks, specifying what to implement, including API signatures, parameter semantics, default values, and exception behavior, while withholding implementation details.
Each subtask is verified by its own test suite and contributes to a weighted overall score, so partial progress is captured as a continuous value rather than a binary outcome.
To broaden coverage, we curate 17 repositories across 5 programming languages, spanning data processing, web frameworks, ORMs, serialization, GUI toolkits, and developer tooling, with no overlap with existing benchmarks.
To ensure that failures reflect genuine capability gaps rather than benchmark artifacts, we combine static validation with attribution-driven rollout-based quality control to separate task-side defects from model-side limitations and iteratively repair confirmed task issues.

We evaluate thirteen frontier models on \MYBENCH~ and observe that no model comes close to solving the benchmark. 
As shown in Figure~\ref{fig:leaderboard}, even the strongest model, Claude-Opus-4.7, resolves only 39.1\% of tasks, while the weakest achieves merely 5.2\%. By comparison, these systems attain 80\%+ scores on SWE-bench Verified~\citep{openai2024sweverified}. 
The Completion Score reveals a consistent pattern: models routinely complete several subtasks before stalling at integration boundaries, offering cleaner separation across capability tiers than binary outcomes alone.

In summary, our key contributions are as follows:
\begin{itemize}[leftmargin=*, itemsep=2pt, topsep=4pt]
  \item We construct \MYBENCH, a benchmark of 115 real open-source version-upgrade tasks across 17 repositories and 5 programming languages, establishing long-horizon multi-target software development as a distinct evaluation setting.
  \item We develop a construction pipeline that transforms real version upgrades into multi-target tasks, and combines static validation with rollout-based quality control to separate task-side defects from genuine model limitations and iteratively repair confirmed issues.
  \item We evaluate thirteen frontier models and find that resolved rates range from 5.2\% to 39.1\%, well below performance on existing bug-fix benchmarks, while Completion Score reveals fine-grained capability differences across domains and difficulty tiers beyond binary resolved metrics.
\end{itemize}
\section{Related Work}

\paragraph{Coding Agents.}
LLM-based coding agents have evolved from single-turn code generation systems to interactive software engineering agents operating in realistic development environments\citep{sapkota2025vibe,dong2025surveycodegenerationllmbased,starace2025paperbench}.
OpenHands~\cite{wang2024openhands} provides an open platform for building generalist software development agents, while Terminus~2~\cite{terminalbench2026} serves as the reference agent implementation within the Harbor framework for autonomous evaluation in sandboxed environments.
Commercial systems such as Claude Code~\cite{anthropic2025claudecode} have further brought agentic coding into mainstream software development workflows.
As these systems become increasingly capable, there is a growing need for benchmarks that better reflect the complexity of real-world software engineering.


\paragraph{Coding Benchmarks for Agents.}
Coding benchmarks for LLM agents have progressively evolved from function-level synthesis to more realistic software engineering tasks. HumanEval~\cite{chen2021humaneval} and MBPP~\cite{austin2021mbpp} focus on function-level code generation. The SWE-bench family~\cite{jimenez2024swebench, openai2024sweverified, deng2025swe} extends evaluation to issue resolution and long-horizon engineering tasks in real-world repositories. Later benchmarks broaden evaluation to feature-oriented development and system-level interaction, including FeatureBench~\cite{zhou2026featurebench} and TerminalBench~\cite{terminalbench2026}. More recent work explores increasingly open-ended and long-horizon software engineering settings. NL2Repo~\cite{ding2025nl2repo} evaluates full repository generation from natural language specifications without requiring agents to evolve existing large-scale codebases, while SWE-EVO~\cite{thai2025swe} studies Python version evolution but derives problem statements directly from release notes without explicit instruction-test alignment validation. Existing benchmarks still primarily evaluate isolated tasks rather than structured long-horizon multi-target software development processes. \MYBENCH~ covers 17 repositories across 5 programming languages, where each instance contains around five structured subtasks together with dedicated instruction-test alignment validation. Table~\ref{tab:comparison} summarizes the key differences among existing benchmarks.

\section{\MYBENCH~}
\label{sec:benchmark}

We describe \MYBENCH~ across three aspects: the task definition and evaluation protocol (Section~\ref{sec:formulation}), dataset statistics (Section~\ref{sec:stats}), and the construction pipeline (Section~\ref{sec:pipeline}).

\subsection{Task Definition}
\label{sec:formulation}

\begin{figure}[t]
  \centering
  \includegraphics[width=\linewidth]{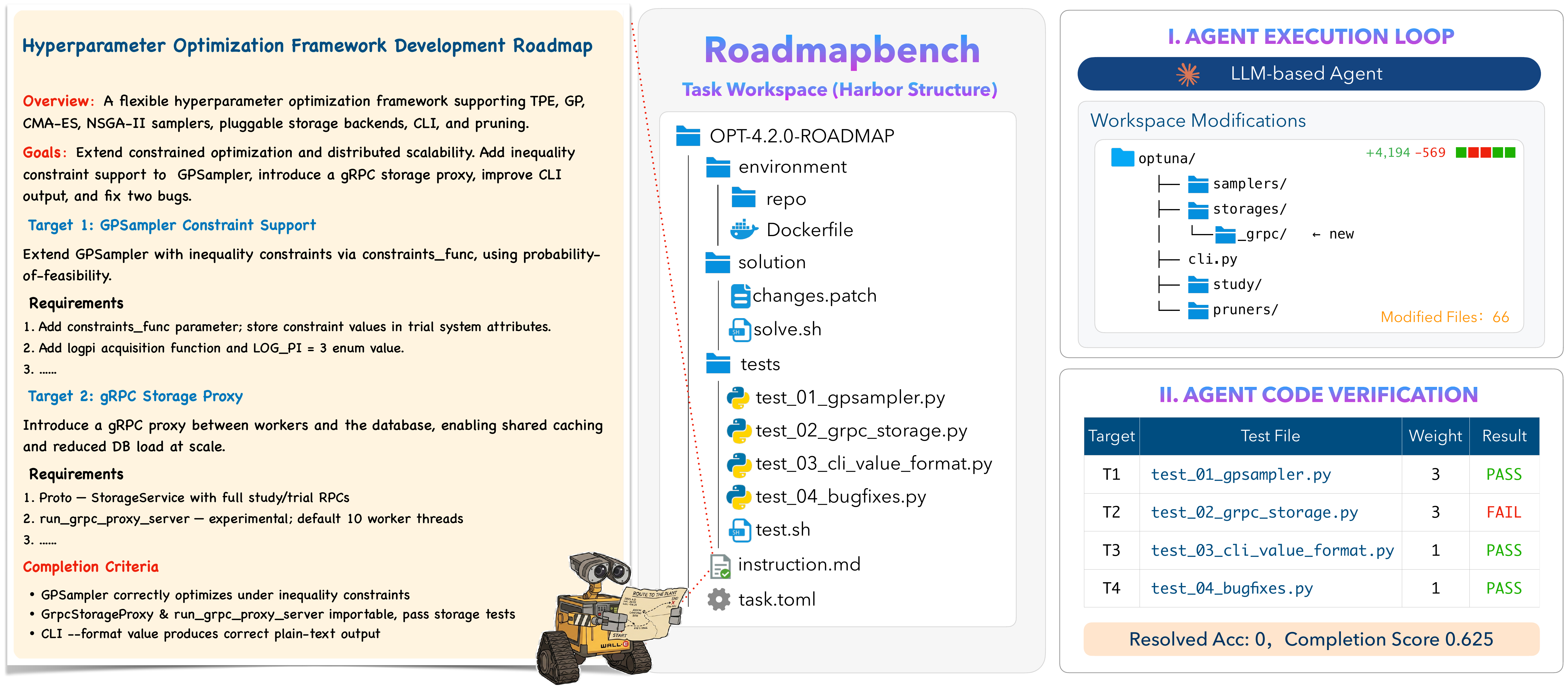}
  \caption{\textbf{Overview of a \MYBENCH~ task.}
  The agent receives a source-version repository snapshot and a roadmap-style instruction, then implements the specified functionality inside a pinned Docker environment.
  Evaluation is performed via weighted subtask-level tests against behaviors introduced in the target version.}
  \label{fig:task_example}
\end{figure}

As illustrated in Figure~\ref{fig:task_example}, each \MYBENCH~ task asks an agent to implement the functionality introduced in a real version upgrade.
The agent operates in a Docker environment with the repository pinned at the source version.
It is given a multi-target roadmap instruction specifying \emph{what} to implement: each target corresponds to a distinct unit of new functionality and describes the expected behavioral requirements.
As in real version upgrades, where multiple substantial changes are coordinated within a single release, the targets collectively capture a unified development objective.
We evaluate each task along two dimensions.
A task is \emph{resolved} if the agent passes all subtasks, providing a binary measure of complete success.
To capture partial progress, we additionally compute a weighted reward: each subtask carries a weight reflecting its implementation complexity, and the reward is the weighted fraction of passed subtasks.

\subsection{Dataset Statistics}
\label{sec:stats}

\begin{figure*}[t]
  \centering
  \includegraphics[width=\linewidth]{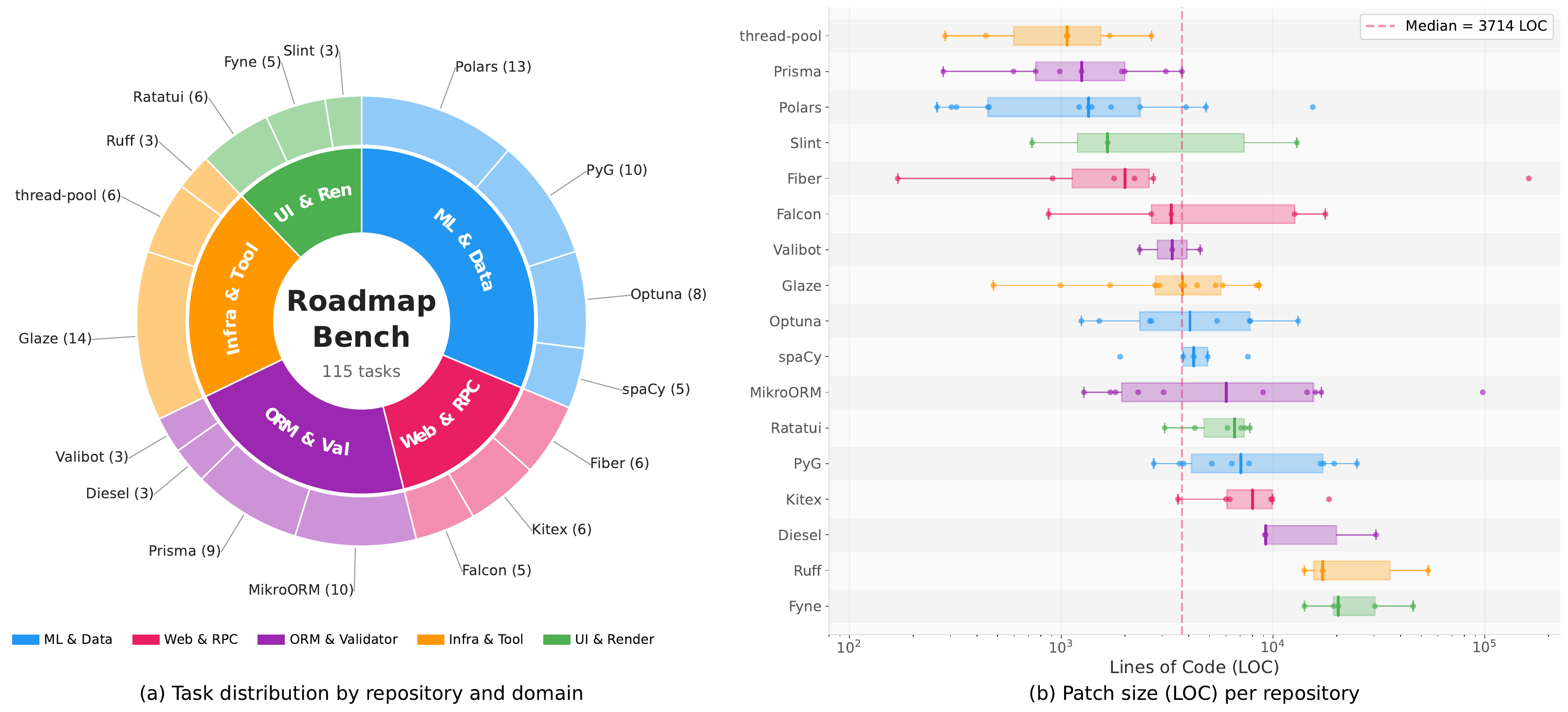}
  \caption{\textbf{Dataset overview of \MYBENCH.}
  (a) Task count per repository (outer ring) grouped by domain (inner ring): ML \& Data (36), Web \& RPC (17), ORM \& Val (25), Infra \& Tool (23), UI \& Ren (14).
  (b) Distribution of ground-truth patch size (lines changed) per repository, where the dashed line marks the overall median of 3{,}714 LOC.}
  \label{fig:stats}
\end{figure*}

The current release contains 115 tasks spanning 17 open-source repositories across five programming languages (see Appendix~\ref{app:task_details} for details).
Oracle patches range from under 300 to over 30{,}000 lines changed, with a median of approximately 3{,}700 lines and 51 files touched.
Subtask counts range from 3 to 12 with a median of 5, confirming that tasks require sustained multi-target engineering rather than single-function edits.
Figure~\ref{fig:stats} shows the task distribution and oracle patch size across repositories. 

\subsection{Data Construction Pipeline}
\label{sec:pipeline}


\begin{figure}[t]
  \centering
  \includegraphics[width=\linewidth]{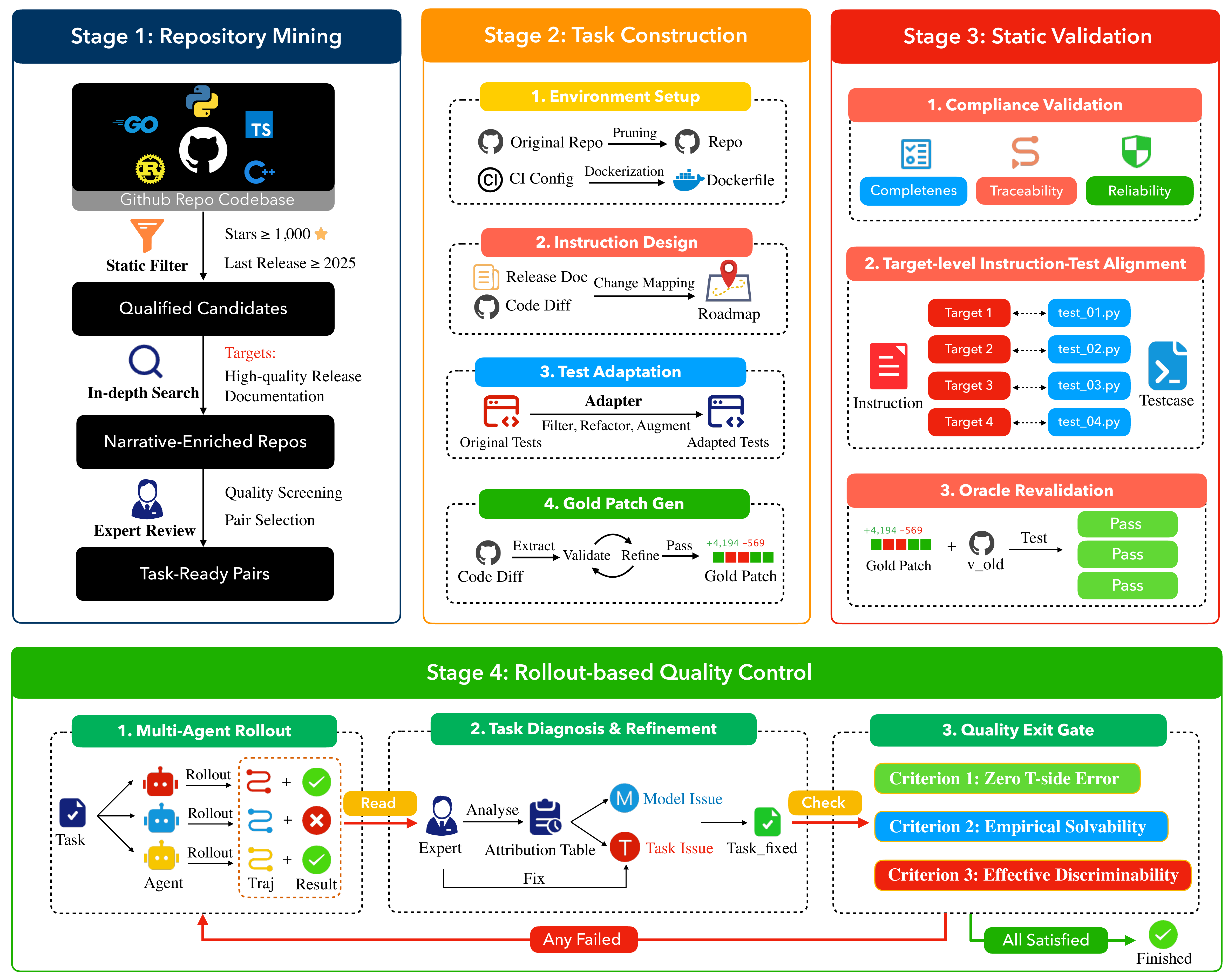}
  \caption{\textbf{\MYBENCH~ construction pipeline.}
  Repository mining selects task-ready version pairs; task construction aligns release narratives with code diffs to create instructions and tests. Static validation and rollout-based quality control repair task-side defects before benchmark inclusion.}
  \label{fig:construction_pipeline}
\end{figure}

The pipeline proceeds in four stages (Figure~\ref{fig:construction_pipeline}): repository mining, task construction, static validation, and rollout-based quality control.


\paragraph{Stage 1: Repository Mining.}
We aggregate repositories from community-curated open-source project lists across five languages and apply a three-stage filter: (1)~a rule-based filter retains repositories with at least 1{,}000 stars, five or more tagged releases, and continued release activity through 2025; (2)~an in-depth search identifies repositories that maintain high-quality release documentation (see examples in Appendix~\ref{app:repo_selection}); (3)~expert review verifies documentation quality and selects consecutive version pairs with sufficient code changes and feature narratives for task construction.
This process yields 17 repositories and 115 version pairs across five languages.\looseness=-1


\paragraph{Stage 2: Task Construction.}
Each task is built in a Docker environment pinned to the source version.
The git history is preserved but all branches and tags beyond the source release are pruned, preventing the agent from inspecting target-version code through version control.
We align source-to-target code diffs with release narratives to identify externally visible behavioral changes and create a multi-target roadmap instruction (\texttt{instruction.md}) specifying what to implement without revealing how.
Tests are adapted from upstream suites to preserve behavioral coverage, and a gold patch is extracted from the code diff, refined against the task environment, and validated until it passes the adapted tests.

\paragraph{Stage 3: Static Validation.}
Each task is statically checked along two dimensions: \emph{compliance}, verifying specification self-containedness, source traceability, and test validity; and \emph{target-level correctness}, ensuring that every tested behavior for each target is specified and no test relies on unstated assumptions (details in Appendix~\ref{app:static_review}).
Confirmed issues are repaired; the oracle patch is then re-run to ensure the fail-to-pass guarantee remains valid.

\paragraph{Stage 4: Rollout-based quality control.}
Agents from three capability tiers attempt each task; failures are attributed to either task-side defects (missing/ambiguous specifications) or genuine model limitations (incorrect design, buggy implementation).
Task-side defects are iteratively repaired and revalidated until cleared. A task is finalized only when it contains no task-side errors, the oracle achieves full reward, and models of different tiers produce distinguishable scores (see Appendix~\ref{app:qc}).

\section{Experiments}

\begin{table*}[t]
  \caption{Main results on \MYBENCH~across 115 tasks, using a single trial per model.
Domain columns report resolved rates (\%). Task counts are ML\,\&\,Data~(36), Web\,\&\,RPC~(17), ORM\,\&\,Val.~(25), Infra.\,\&\,Tool.~(23), and UI\,\&\,Ren.~(14).
\textbf{Bold} and \underline{underline} denote the best and second-best domain-level results within each scaffold.
}
  \label{tab:main}
  \centering
  \setlength{\tabcolsep}{4pt}
  \renewcommand{\arraystretch}{1.15}
  \scalebox{0.9}{
  \begin{tabular}{@{}l cccc @{\hskip 8pt} ccccc@{}}
    \toprule
    & \multicolumn{4}{c}{\textbf{Overall}} & \multicolumn{5}{c}{\textbf{Resolved Rate by Domain (\%)}} \\
    \cmidrule(lr){2-5} \cmidrule(lr){6-10}
    \textbf{Model} &
    \textbf{\shortstack{Resolved\\(\%)}} &
    \textbf{\shortstack{Completion\\Score}} &
    \textbf{\shortstack{Avg.\\Turns}} &
    \textbf{\shortstack{Output\\Tok.\,(K)}} &
    \textbf{\shortstack{ML\,\&\\Data}} &
    \textbf{\shortstack{Web\,\&\\RPC}} &
    \textbf{\shortstack{ORM\,\&\\Val.}} &
    \textbf{\shortstack{Infra.\,\&\\Tool.}} &
    \textbf{\shortstack{UI\,\&\\Ren.}} \\
    \midrule
    \multicolumn{10}{c}{\cellcolor{blue!6}\textbf{\textsc{OpenHands}}} \\
    \midrule
    Claude-Opus-4.7   & \textbf{39.1} & \textbf{0.692} & 140.2 & 44 & \textbf{30.6} & \textbf{41.2} & \textbf{32.0} & \textbf{43.5} & \textbf{64.3} \\
    Claude-Opus-4.6   & \underline{32.2} & \underline{0.627} & 140.7 & 42 & 25.0 & \underline{29.4} & \textbf{32.0} & 30.4 & \underline{57.1} \\
    GPT-5.4           & 29.6 & 0.497 & 170.7 & 93 & \underline{27.8} & 17.6 & 20.0 & \underline{39.1} & 50.0 \\
    Gemini-3.1-Pro    & 20.9 & 0.439 & 133.4 & 26 &  8.3 & 23.5 & 24.0 & 26.1 & 35.7 \\
    DeepSeek-V4-Pro   & 18.3 & 0.486 & 140.2 & 64 &  8.3 & 17.6 & 24.0 & 17.4 & 35.7 \\
    GLM-5.1           & 18.3 & 0.453 & 163.2 & 38 &  8.3 & 11.8 & \underline{28.0} & 26.1 & 21.4 \\
    Kimi-K2.6         & 14.8 & 0.432 & 158.9 & 76 &  5.6 &  5.9 & 20.0 & 21.7 & 28.6 \\
    Mimo-V2.5-Pro     & 13.9 & 0.440 & 155.5 & 66 &  8.3 & 11.8 & 12.0 & 21.7 & 21.4 \\
    Qwen3.6-Plus      & 12.2 & 0.424 & 150.3 & 47 &  5.6 &  5.9 & 12.0 & 21.7 & 21.4 \\
    Kimi-K2.5         & 11.3 & 0.378 & 110.3 & 29 &  0.0 &  5.9 & 12.0 & 17.4 & 35.7 \\
    MiniMax-M2.7      & 10.4 & 0.332 & 123.5 & 38 &  5.6 &  0.0 &  8.0 & 26.1 & 14.3 \\
    Qwen3.5-397B      &  9.6 & 0.383 & 110.5 & 35 &  0.0 & 11.8 & 12.0 & 13.0 & 21.4 \\
    Seed-2.0-Pro      &  5.2 & 0.177 &  40.1 &  9 &  0.0 &  5.9 &  8.0 &  4.3 & 14.3 \\
    \midrule
    \multicolumn{10}{c}{\cellcolor{blue!6}\textbf{\textsc{Terminus~2}}} \\
    \midrule
    Claude-Opus-4.7   & \textbf{38.3} & \textbf{0.681} &  59.2 & 22 & \textbf{27.8} & \textbf{41.2} & \textbf{36.0} & \textbf{43.5} & \textbf{57.1} \\
    Claude-Opus-4.6   & \underline{31.3} & \underline{0.666} &  82.7 & 43 & \underline{19.4} & \underline{23.5} & \textbf{36.0} & \underline{39.1} & \underline{50.0} \\
    GLM-5.1           & 20.9 & 0.512 &  93.8 & 57 & 11.1 & 11.8 & \underline{32.0} & 21.7 & 35.7 \\
    Qwen3.6-Plus      & 16.5 & 0.508 & 129.6 & 64 &  8.3 & 11.8 & 16.0 & 26.1 & 28.6 \\
    Kimi-K2.6         & 15.7 & 0.409 & 111.7 & 53 &  5.6 & 11.8 & 28.0 & 17.4 & 21.4 \\
    DeepSeek-V4-Pro   & 10.4 & 0.395 & 149.2 & 80 &  2.8 &  5.9 & 12.0 & 21.7 & 14.3 \\
    Mimo-V2.5-Pro     & 10.4 & 0.344 & 113.7 & 155 &  2.8 & 17.6 &  8.0 & 13.0 & 21.4 \\
    Qwen3.5-397B      & 10.4 & 0.337 &  90.1 & 43 &  2.8 &  5.9 & 12.0 & 21.7 & 14.3 \\
    Kimi-K2.5         &  7.8 & 0.360 &  90.2 & 33 &  0.0 &  0.0 & 16.0 & 17.4 &  7.1 \\
    MiniMax-M2.7      &  4.3 & 0.279 & 126.2 & 41 &  0.0 &  0.0 &  4.0 & 13.0 &  7.1 \\
    Seed-2.0-Pro      &  2.6 & 0.135 &  55.9 & 20 &  0.0 &  0.0 & 12.0 &  0.0 &  0.0 \\
    \bottomrule
  \end{tabular}}
\end{table*}

\subsection{Evaluation Setup}
\label{sec:eval_setup}

\paragraph{Models.}
We evaluate thirteen frontier models: Claude-Opus-4.7~\cite{anthropic2025claude47}, Claude-Opus-4.6~\cite{anthropic2025claude46}, GPT-5.4~\cite{openai2026gpt54}, Gemini-3.1-Pro~\cite{googledeepmind2026gemini31pro}, DeepSeek-V4-Pro~\cite{deepseekai2026deepseekv4}, GLM-5.1~\cite{glm5team2026}, Kimi-K2.6~\cite{team2026kimik26}, Mimo-V2.5-Pro~\cite{xiaomi2026mimo}, Qwen3.6-Plus~\cite{qwen36plus}, Kimi-K2.5~\cite{team2026kimi}, MiniMax-M2.7~\cite{minimax2026m27}, Qwen3.5-397B~\cite{qwen35397b}, and Seed-2.0-Pro~\cite{bytedance2026seed2}.
These models span multiple commercial API providers and cover a wide range of current capability tiers.

\paragraph{Agent scaffold.}
All tasks are packaged as Harbor~\cite{harbor2026} environments and can be evaluated with any Harbor-compatible agent.
We use OpenHands~\cite{wang2024openhands} as the primary scaffold for all thirteen models.
Each rollout runs inside a pinned Docker environment rooted at the source version.
The agent may inspect and modify the repository but has no access to target-version code, test files, or the oracle patch.
Future branches and upstream repository access are blocked to prevent information leakage.
As an ablation, we additionally evaluate a subset of models under Terminus~2, the reference agent implementation of Harbor.
Terminus~2 is designed as a neutral testing platform that runs fully autonomously in sandboxed environments, making it well suited for measuring scaffold sensitivity independent of any production-oriented design choices in OpenHands.

\paragraph{Inference configuration.}
Each task is allocated a 2-hour wall-clock budget.
All models are evaluated with extended thinking enabled.
For models that support configurable reasoning depth, we set reasoning effort to \texttt{high} for GPT-5.4, Gemini-3.1-Pro, DeepSeek-V4-Pro, and Seed-2.0-Pro, and \texttt{xhigh} for Claude-Opus-4.7; the remaining models use their default thinking mode.

\paragraph{Metrics.}
For task $t$ with $K_t$ subtasks, each subtask $k$ carries a weight $w_{t,k}$ reflecting its relative complexity and yields a binary pass/fail result $r_{t,k} \in \{0,1\}$.
We define the per-task weighted reward as
\[
  s_t \;=\; \frac{\sum_{k=1}^{K_t} w_{t,k} \cdot r_{t,k}}{\sum_{k=1}^{K_t} w_{t,k}},
\]
We report two primary metrics over $N$ tasks: one for full task completion and one for partial progress.
\emph{Resolved rate} is the fraction of fully completed tasks: $\operatorname{RR} = \frac{1}{N}\sum_{t} \mathbf{1}[s_t = 1]$.
\emph{Completion Score} averages $s_t$ to credit partial completions: $\operatorname{CS} = \frac{1}{N}\sum_{t} s_t$.
We also report \emph{Avg.\ turns}, the mean number of agent turns per task, and \emph{Output Tok.}, the average output tokens generated per task (in thousands), as indicators of interaction cost and computational effort.

\subsection{Main Results}
\label{sec:results}

Table~\ref{tab:main} reports resolved rate, Completion Score, average turns, and per-domain resolved rates for thirteen frontier models under OpenHands, with Terminus~2 results for a subset of models alongside for comparison.
A detailed scaffold sensitivity analysis is provided in \S\ref{app:scaffold}.

\paragraph{Overall performance.}
Current frontier models remain far from solving \MYBENCH.
Under OpenHands, Claude-Opus-4.7 achieves the highest resolved rate at 39.1\%, followed by Claude-Opus-4.6 at 32.2\% and GPT-5.4 at 29.6\%.
The remaining ten models range from 5.2\% to 20.9\%, indicating a substantial gap between the strongest models and the rest.
Completion Score highlights that partial progress is common.
It is consistently higher than resolved rate across models, showing that agents often complete some roadmap targets before failing to solve the full task.
For example, Claude-Opus-4.6 resolves 32.2\% of tasks but obtains a Completion Score of 0.627, while Seed-2.0-Pro resolves 5.2\% yet reaches 0.177.
This suggests that failures often occur after partial progress, when agents stall on later targets, integration, or correctness.

\paragraph{Domain difficulty.}
Performance varies substantially across domains.
ML \& Data is the most challenging: six of thirteen models resolve no tasks, and only the top three exceed 8\%.
ORM \& Validation is relatively more tractable, likely due to the structured nature of schema migration and validation APIs.
UI \& Rendering shows the sharpest separation across capability tiers, with Claude-Opus-4.7 reaching 64.3\% and Claude-Opus-4.6 reaching 57.1\%, while weaker models remain much lower.
Web \& RPC and Infra.\ \& Tooling fall between these extremes, reflecting intermediate levels of domain structure and integration complexity.

\begin{figure}[t]
  \centering
  \includegraphics[width=0.98\linewidth]{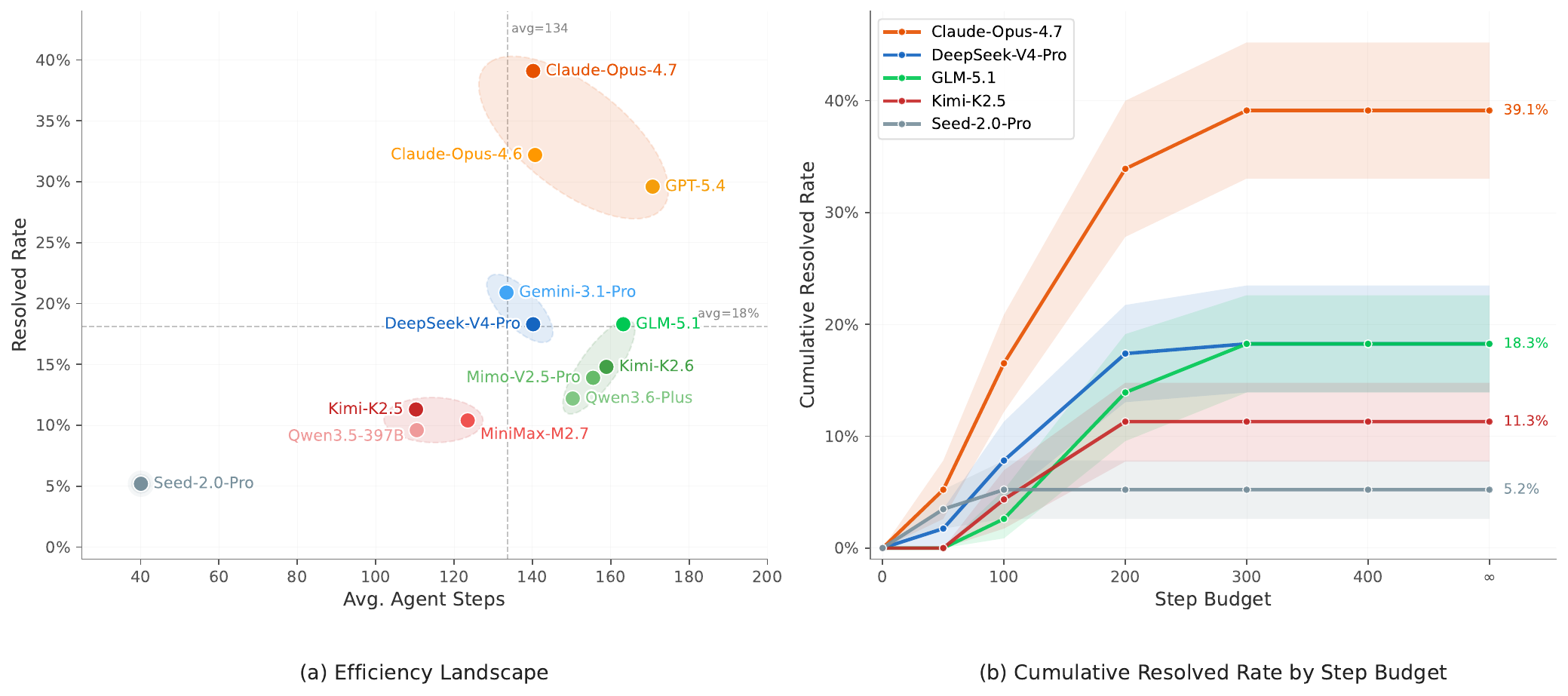}
  \caption{\textbf{Efficiency and step-budget analysis.}
(a) Efficiency landscape of resolved rate versus average agent steps.
Dashed lines mark fleet means, and shaded ellipses indicate performance tiers.
(b) Cumulative resolved rate under increasing per-task step budgets, showing how models convert additional compute into task resolution.}
  \label{fig:behavior}
\end{figure}
\section{Analysis}
\label{sec:analysis}

We decompose performance along six aspects: \textbf{Step Efficiency and Compute Scaling} (\S\ref{sec:step_efficiency}), capturing how much trajectory budget is consumed per resolved task; \textbf{Tool Composition and Usage Distribution} (\S\ref{sec:tool_composition}), capturing how that budget is allocated across different intents; \textbf{Task Complexity and Performance} (\S\ref{app:complexity}), examining how complexity affects resolution; \textbf{Scaffold Sensitivity} (\S\ref{app:scaffold}), comparing agent frameworks; \textbf{Target-Level Analysis} (\S\ref{sec:subtask_analysis}), stratifying by change type and difficulty; and \textbf{Failure Mode Analysis} (\S\ref{sec:failure_mode}), characterizing where unsuccessful trajectories break down.

\subsection{Step Efficiency and Compute Scaling}
\label{sec:step_efficiency}
To characterize behavioral patterns and step efficiency across models, Figure~\ref{fig:behavior}(a) plots average agent steps against resolved rate, with dashed lines marking the fleet averages of 134 steps and 18\%.
The models separate into distinct regimes.
Frontier models, including Claude-Opus-4.7, Claude-Opus-4.6, and GPT-5.4, achieve 30\% to 39\% resolved rates with moderate to high step budgets.
By contrast, models such as GLM-5.1 and Kimi-K2.6 consume comparable or larger budgets but remain near the mid-performance region, indicating lower step efficiency.
This contrast is particularly clear for Claude-Opus-4.7 and GLM-5.1, which use similar average budgets, 140 and 163 steps respectively, yet differ by more than 20 percentage points in resolved rate.
Seed-2.0-Pro appears as a low-compute, low-performance outlier, suggesting premature termination or limited repository interaction.

Figure~\ref{fig:behavior}(b) shows the cumulative resolved rate as the per-task step budget increases.
Most models saturate within the first 200 steps, after which additional budget provides limited gains.
The strongest model, Claude-Opus-4.7, is the main exception, continuing to improve beyond this point and reaching 39.1\% at the full budget.
Among mid-tier models, DeepSeek-V4-Pro and GLM-5.1 reach similar final resolved rates but follow different scaling trajectories.
DeepSeek-V4-Pro plateaus earlier, indicating higher step efficiency, whereas GLM-5.1 requires a larger budget to approach the same level.
These trends suggest that additional steps are beneficial only when models can effectively convert longer trajectories into successful edits.

\begin{figure}[t]
  \centering
  \includegraphics[width=\linewidth]{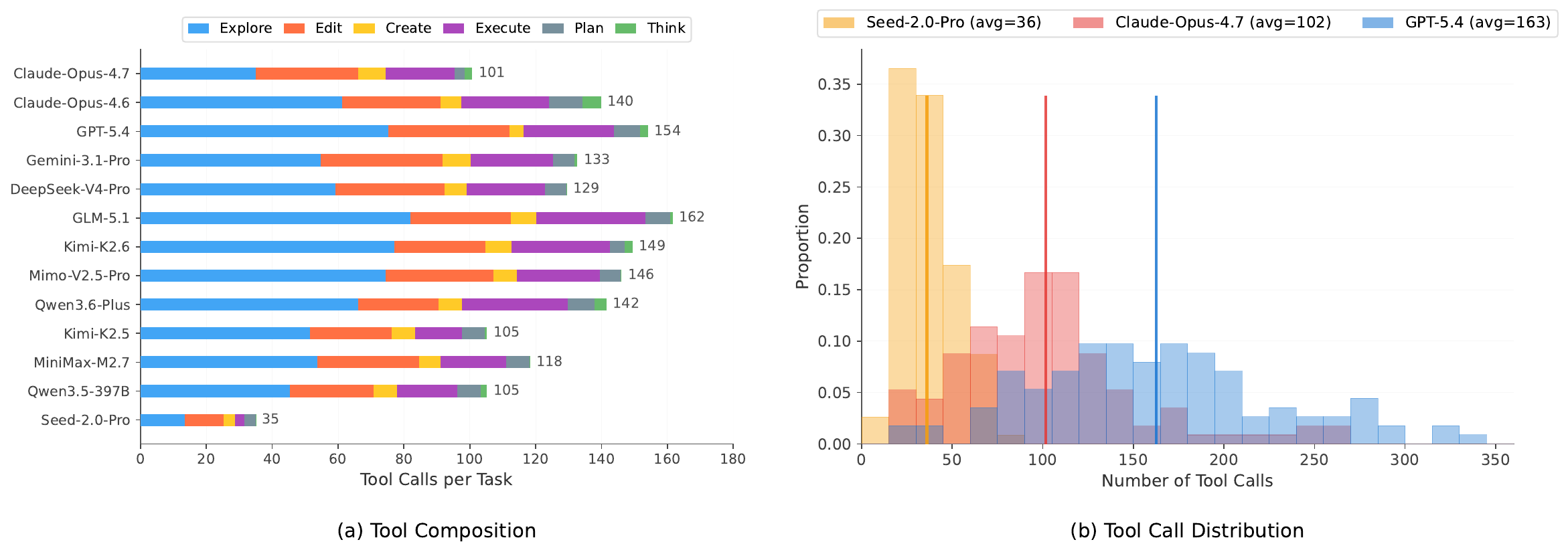}
  \caption{\textbf{Tool usage analysis.} (a) Tool composition by model, decomposed into six intent categories and sorted by resolved rate. (b) Distribution of per-task tool call counts for three representative models spanning the full performance range: Seed-2.0-Pro (5\%), Claude-Opus-4.7 (39\%), and GPT-5.4 (30\%). Vertical lines indicate mean values.}
  \label{fig:tool_combined}
\end{figure}

\subsection{Tool Composition and Usage Distribution}
\label{sec:tool_composition}
We classify each tool invocation into six intent-based categories derived from the OpenHands agent's action space.
\emph{Explore} encompasses file viewing (\texttt{str\_replace\_\allowbreak{}editor\allowbreak{} view}) and shell-based search or inspection commands (e.g., \texttt{grep}, \texttt{find}, \texttt{cat});
\emph{Edit} covers in-place code modifications (\texttt{str\_replace\_\allowbreak{}editor\allowbreak{} str\_replace}) and shell editing commands;
\emph{Create} captures new file creation (\texttt{str\_replace\_\allowbreak{}editor\allowbreak{} create/\allowbreak{}insert});
\emph{Execute} includes compilation, testing, dependency installation, and other shell executions;
\emph{Plan} corresponds to explicit task planning via the built-in task tracker; and
\emph{Think} represents deliberate reasoning steps.
Terminal actions (e.g., task completion) and tool misuse are excluded.

As shown in Figure~\ref{fig:tool_combined}(a),
\emph{Explore}, \emph{Edit}, and \emph{Execute} dominate tool usage across all models, corresponding to repository inspection, code modification, and validation.
The main difference across models is not the amount of tool use, but how tool calls are allocated across the development process.
Claude-Opus-4.7 achieves the highest resolved rate with only 101 tool calls per task on average and the lowest \emph{Explore} ratio at 35\%.
In contrast, GLM-5.1 and Kimi-K2.6 use substantially more tool calls, but spend over half of them on exploration.
This suggests that strong models localize relevant code more efficiently and shift earlier from exploration to targeted editing and execution-based validation.
Explicit \emph{Plan} and \emph{Think} calls remain sparse for most models, indicating that the observed trajectories are driven mainly by iterative exploration, editing, and execution rather than dedicated reasoning-oriented tool actions.

Figure~\ref{fig:tool_combined}(b) compares the per-task tool call distributions of three representative models.
Seed-2.0-Pro uses only 36 tool calls on average and obtains a low resolved rate, suggesting insufficient repository interaction.
GPT-5.4 uses 163 tool calls on average, indicating much longer trajectories.
Claude-Opus-4.7 reaches the best resolved rate with an intermediate average of 102 tool calls.
Overall, these results indicate that task success is better characterized by the allocation of tool use across exploration, editing, planning, and execution than by raw tool-call volume alone.

\subsection{Task Complexity and Performance}
\label{app:complexity}

\begin{figure}[t]
  \centering
  \includegraphics[width=\linewidth]{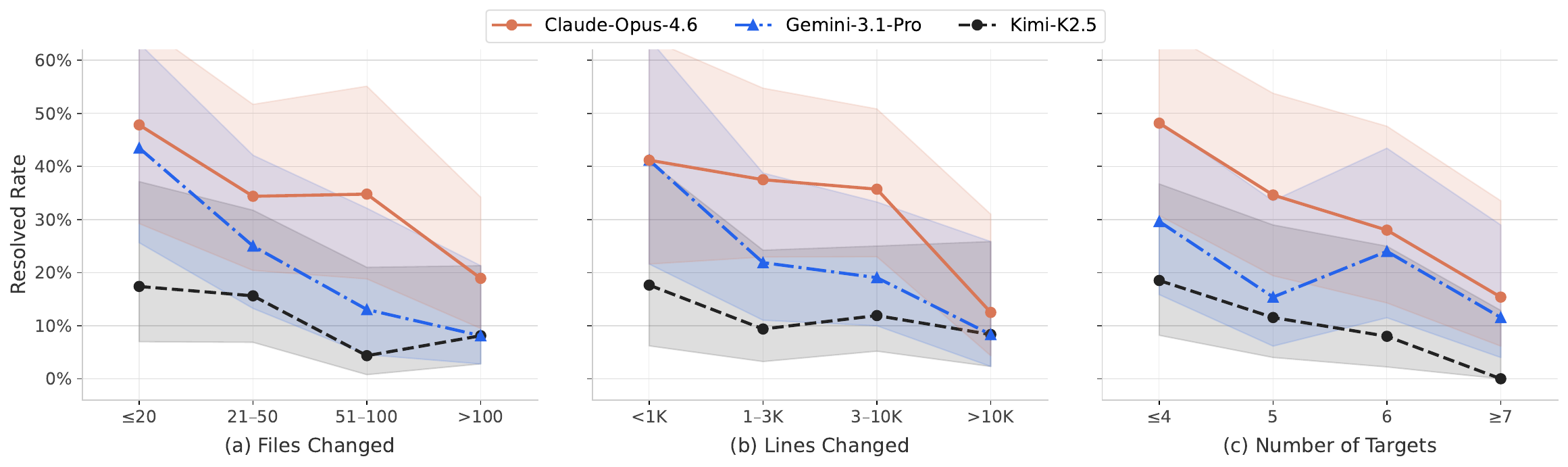}
  \caption{\textbf{Resolved rate vs.\ three task complexity proxies} (binned rate $\pm$ 95\% Wilson CI). (a) Files changed, (b) lines changed, and (c) number of targets are all strong predictors of task difficulty, with monotonically decreasing resolved rates as complexity increases.}
  \label{fig:complexity}
\end{figure}

Resolved rate declines consistently as task complexity increases across all three structural proxies (Figure~\ref{fig:complexity}).
\emph{Files changed} (a) shows the clearest model separation: stronger models hold up longer as file count grows, while weaker models fall off early, with Gemini dropping from 43\% to 8\% across the full range---a steeper decline than Claude's 48\% to 19\%.
\emph{Code volume} (b) reveals a more nuanced pattern: on simpler tasks (under 1K lines), Claude and Gemini start at similar levels ($\sim$41\%), but Claude maintains a clear advantage through mid-range complexity while Gemini drops sharply in the intermediate bins; at the hardest end ({>}10K lines), both converge near the floor, suggesting extreme complexity is a ceiling even for the strongest models.
\emph{Subtask count} (c) amplifies this dynamic most dramatically: Kimi-K2.5 collapses to 0\% at 7 or more subtasks while Claude still resolves 15\%, making it the sharpest discriminator among the three proxies.
Together, these results confirm that structural complexity is an effective performance discriminator, with the sharpest separation occurring in the mid-range where model capabilities diverge most.

\subsection{Scaffold Sensitivity}
\label{app:scaffold}

Performance varies across scaffolds for most models, but the direction and magnitude differ by capability tier.
Three patterns emerge from Table~\ref{tab:main}.

\paragraph{Top models are scaffold-robust.}
Claude-Opus-4.6 achieves 31.3\% on Terminus~2 and 32.2\% on OpenHands, a difference of 0.9 percentage points.
Mid- and lower-tier models show larger swings of 3 to 10 percentage points across scaffolds.

\paragraph{OpenHands yields higher performance for most models.}
The majority of evaluated models perform better under OpenHands.
The gains are largest for DeepSeek-V4-Pro (+7.9~pp) and MiniMax-M2.7 (+6.1~pp).
OpenHands provides explicitly typed tool schemas with clear argument names, which reduces the effort required to select and format each tool call correctly.

\paragraph{Two models perform better on Terminus~2.}
GLM-5.1 and Qwen3.6-Plus are the only exceptions, with resolved rates 2.6~pp and 4.3~pp higher on Terminus~2.
Terminus~2 requires the agent to batch multiple commands into a single structured JSON response per turn, a format these two models handle more effectively than the one-action-per-turn interface of OpenHands.

\subsection{Target-Level Analysis}
\label{sec:subtask_analysis}

We classify subtasks into five change types: Component Creation, Feature Addition, Feature Enhancement, Behavior Change, and Bug Fix.
A clear difficulty gradient emerges: average pass rate rises from 36\% (Component Creation) to 64\% (Bug Fix), confirming that designing new abstractions and multi-file coordination is substantially harder than locating and correcting specific defects.

\begin{figure}[t]
  \centering
  \includegraphics[width=\linewidth]{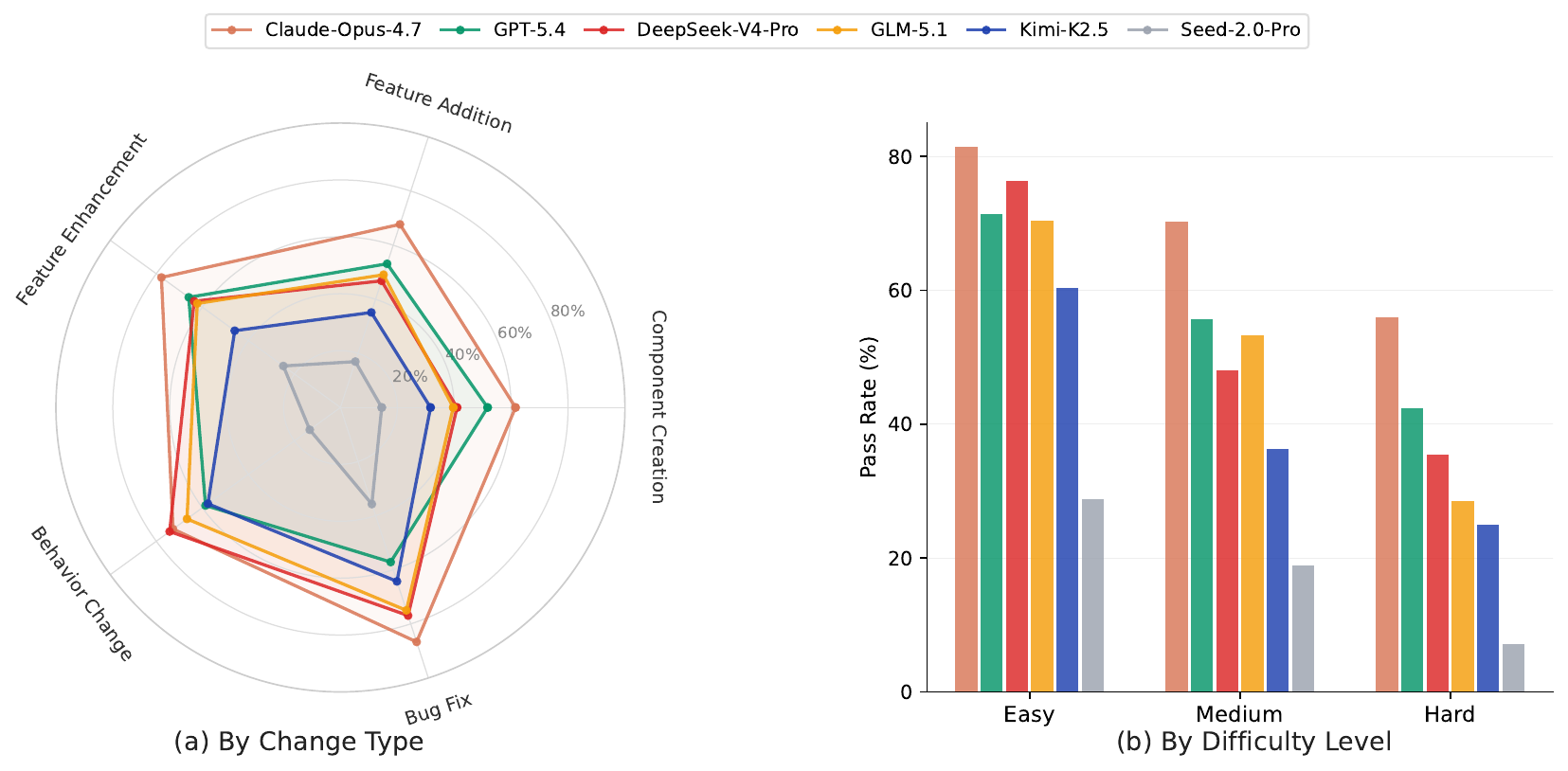}
  \caption{\textbf{Subtask pass rate for six representative models.} (a) By change type. (b) By difficulty level.}
  \label{fig:subtask_radar}
\end{figure}

Figure~\ref{fig:subtask_radar} breaks down performance by change type and difficulty level across six representative models.
Panel (a) reveals that the gap between strong and weak models is most pronounced on Component Creation and Feature Addition, where Claude maintains over 50\% while Seed-2.0-Pro drops below 25\%.
Panel (b) shows that on Hard subtasks, Claude maintains 53\% while DeepSeek drops to 43\% and Seed-2.0-Pro to 16\%, confirming that difficulty amplifies inter-model gaps.

\subsection{Failure Mode Analysis}
\label{sec:failure_mode}

We perform root-cause analysis on 3{,}603 failed subtasks across thirteen models using Claude-Sonnet-4.6 as an agentic classifier.
We categorize failures into five types.
\texttt{Implementation Error} refers to code that compiles but exhibits incorrect behavior.
\texttt{Build Error} denotes solutions that fail to compile or link.
\texttt{Missing Implementation} captures cases where required functionality is absent.
\texttt{Interface Mismatch} covers incorrect API signatures or export paths.
\texttt{Agent Failure} refers to cases where the agent abandons the task or exhausts its budget.

\begin{figure}[t]
  \centering
  \includegraphics[width=\linewidth]{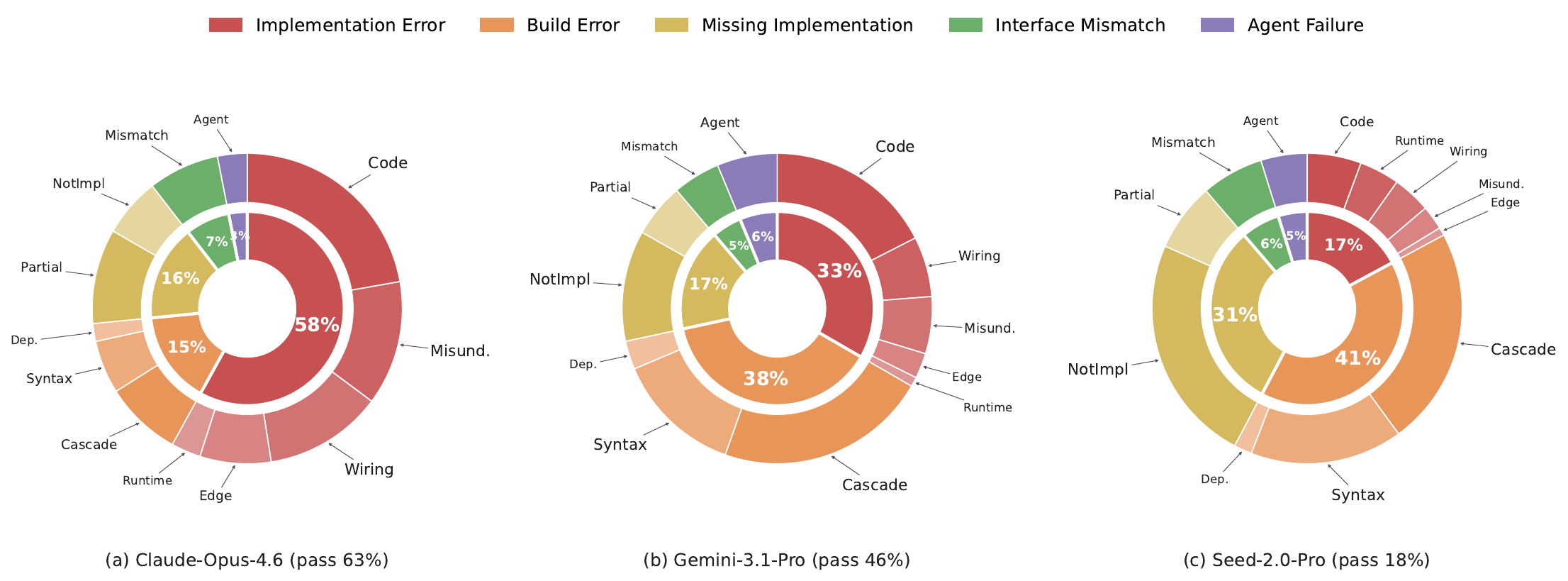}
  \caption{\textbf{Error distribution for three representative models.} Inner ring: category proportions; outer ring: sub-type breakdown. The dominant failure mode shifts from Implementation Error (strong models) to Build Error (weak models).}
  \label{fig:error_donut}
\end{figure}

Figure~\ref{fig:error_donut} reveals a capability-dependent shift in failure modes.
Higher-performing models are less often blocked by construction-level errors such as build failures or missing functionality; instead, their failures concentrate on implementation-level correctness.
For Claude-Opus-4.6, 58\% of failures are \texttt{Implementation Errors}, indicating that the model usually produces complete and buildable code but still fails on behavioral correctness.
These errors are further dominated by \texttt{Code Defect}, \texttt{Misunderstanding}, and \texttt{Wiring Error}, suggesting that the frontier bottleneck lies in execution precision, including subtle logic mistakes, requirement misinterpretation, and component integration.
Gemini-3.1-Pro presents a transitional profile, with \texttt{Build Error} and \texttt{Implementation Error} contributing comparable shares, 38\% and 33\%, respectively.
Seed-2.0-Pro is dominated by earlier construction failures, with \texttt{Build Error} and \texttt{Missing Implementation} accounting for 41\% and 31\% of failures.
This pattern indicates that, as model capability decreases, the primary bottleneck shifts from implementing the correct behavior to producing complete and buildable code.

\section{Conclusion}

\MYBENCH~ introduces a new evaluation axis for coding agents: multi-target, long-horizon software development across real version upgrades.
Each task requires agents to interpret roadmap specifications, coordinate multi-file changes, and implement coherent feature sets.
Across 115 tasks from 17 repositories and 5 programming languages, current models remain far from solving this setting. Under OpenHands, Claude-Opus-4.7 resolves only 39.1\% of tasks, while Seed-2.0-Pro resolves 5.2\%.
Completion Score shows that partial progress is common: agents often complete a subset of roadmap targets before failing on integration, correctness, or construction-level reliability.
Domain-level results further show uneven difficulty, with ML \& Data being the most challenging, ORM \& Validation relatively more tractable, and UI \& Rendering exhibiting a large gap between frontier and weaker models.
Our analysis indicates that stronger models more efficiently localize relevant code and convert exploration into targeted edits, whereas weaker models often fail earlier through build errors or missing implementations.
These results position \MYBENCH~ as a diagnostic benchmark for measuring sustained software development capability beyond isolated issue resolution.

\bibliography{custom}
\bibliographystyle{unsrtnat}

\clearpage
\appendix
\clearpage
\appendix
\onecolumn

{\large\bfseries Appendix}
\vspace{0.8em}

\section{Limitations}
\label{app:limitations}

We acknowledge several limitations of this work.
Our evaluation employs two agent scaffolds (OpenHands and Terminus~2). Agent performance is sensitive to scaffold design choices, and results under other frameworks may differ.
Evaluation relies on test suites that verify behavioral correctness but do not assess code quality, maintainability, or adherence to idiomatic patterns. Future work could incorporate multi-dimensional metrics for a more holistic assessment.
Although ROADMAPBENCH spans five programming languages and multiple software domains, it still covers only a limited subset of real-world development ecosystems. Future extensions could incorporate additional languages, frameworks, and application settings.

\section{Ethics Statement}
\label{app:ethics}

This research conforms to the Code of Ethics.
All benchmark tasks are derived from publicly available open-source repositories.
No private or proprietary code is included.
Repository identities and version numbers are anonymized in the task instructions to prevent information leakage during evaluation, and no personally identifiable information is collected or used.
Human annotators involved in quality control are co-authors of this work and participated voluntarily.

\section{Broader Impacts}
\label{app:broader_impacts}

\MYBENCH~ is designed to measure and advance the capability of coding agents on realistic software engineering tasks.
On the positive side, improved coding agents can increase developer productivity, lower barriers to software development, and accelerate open-source contributions.
On the negative side, more capable coding agents could potentially be misused to generate malicious code or exploit vulnerabilities at scale.
However, our benchmark evaluates agents on constructive software development tasks (implementing features from public roadmaps) rather than adversarial capabilities. We do not release any model weights or fine-tuning recipes.
We believe the diagnostic value of understanding where current agents fail outweighs the marginal risk, as the benchmark primarily reveals limitations rather than enabling new harmful capabilities.

\section{Human Evaluation}
\label{sec:human_eval}

We conduct human evaluation as part of the task construction and quality-control pipeline.
All annotators are Ph.D. students with computer science backgrounds and relevant experience in software engineering.
They participated in constructing coding tasks, reviewing generated instructions, and repairing task-side defects identified during validation. The annotators were compensated above the local minimum hourly wage.
\clearpage
\section{Task Details}
\label{app:task_details}

This appendix provides detailed statistics that supplement the dataset overview in Section~\ref{sec:stats}.

\subsection{Repository and Language Coverage}

Table~\ref{tab:coverage_detail} summarizes the repository coverage of our benchmark across five programming languages and diverse software domains.

\begin{table}[h]
  \setlength{\tabcolsep}{8pt}
  \caption{Repository coverage by language, domain, task count, and median oracle-patch complexity.}
  \vspace{5pt}
  \label{tab:coverage_detail}
  \centering
  \small
  \begin{tabular}{llccccc}
    \toprule
    \textbf{Language} & \textbf{Repository} & \textbf{Tasks} & \textbf{Domain} & \textbf{Med. Lines} & \textbf{Med. Files} & \textbf{Med. Subtasks} \\
    \midrule
    \multirow{5}{*}{Python}
      & Polars      & 13 & ML \& Data    & 1{,}346  & 42  & 5 \\
      & PyG         & 10 & ML \& Data    & 7{,}044  & 140 & 6 \\
      & Optuna      & 8  & ML \& Data    & 4{,}054  & 82  & 5 \\
      & spaCy       & 5  & ML \& Data    & 4{,}226  & 135 & 6 \\
      & Falcon      & 5  & Web \& RPC    & 3{,}311  & 44  & 6 \\
    \midrule
    \multirow{3}{*}{TypeScript}
      & MikroORM    & 10 & ORM \& Val    & 6{,}006  & 122 & 6 \\
      & Prisma      & 9  & ORM \& Val    & 1{,}246  & 30  & 4 \\
      & Valibot     & 3  & ORM \& Val    & 3{,}341  & 56  & 5 \\
    \midrule
    \multirow{2}{*}{C++}
      & Glaze       & 14 & Infra \& Tool & 3{,}745  & 29  & 5 \\
      & thread-pool & 6  & Infra \& Tool & 1{,}065  & 2   & 4 \\
    \midrule
    \multirow{3}{*}{Go}
      & Fiber       & 6  & Web \& RPC    & 1{,}997  & 25  & 6 \\
      & Kitex       & 6  & Web \& RPC    & 8{,}018  & 149 & 5 \\
      & Fyne        & 5  & UI \& Ren     & 20{,}339 & 876 & 7 \\
    \midrule
    \multirow{4}{*}{Rust}
      & Ratatui     & 6  & UI \& Ren     & 6{,}575  & 44  & 6 \\
      & Diesel      & 3  & ORM \& Val    & 9{,}233  & 169 & 4 \\
      & Slint       & 3  & UI \& Ren     & 1{,}656  & 29  & 4 \\
      & Ruff        & 3  & Infra \& Tool & 17{,}130 & 357 & 6 \\
    \midrule
    \multicolumn{2}{l}{\textbf{Overall (17 repos)}} & \textbf{115} & & \textbf{3{,}714} & \textbf{51} & \textbf{5} \\
    \bottomrule
  \end{tabular}
\end{table}

\subsection{Task Complexity Distribution}

\begin{figure}[h]
  \centering
  \includegraphics[width=\linewidth]{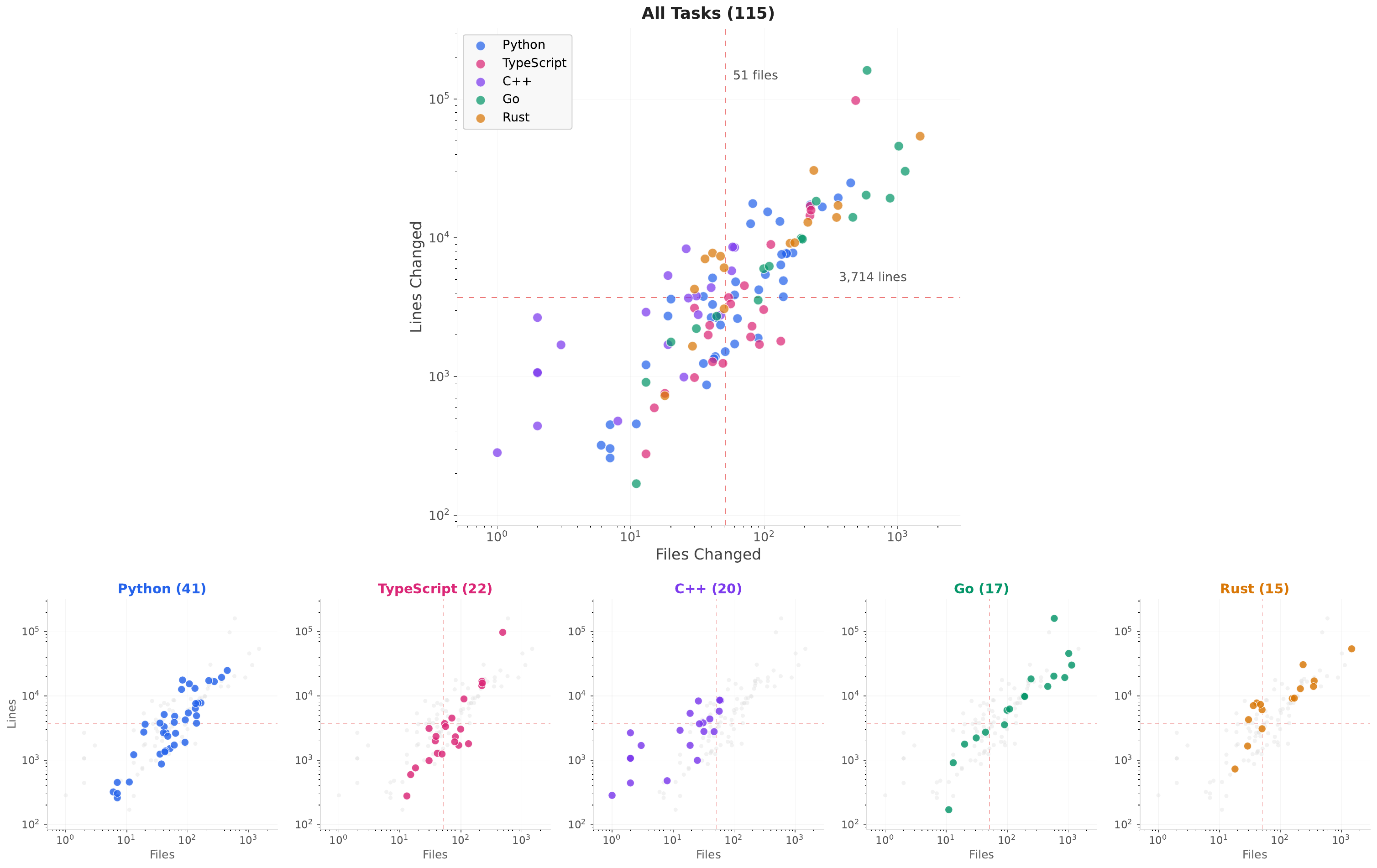}
  \caption{\textbf{Task complexity overview and per-language breakdown} (log-log scale).
  (a) All 115 tasks colored by language, with dashed lines at the benchmark medians (3{,}714 lines, 51 files).
  (b) Per-language panels: each language highlighted against the full benchmark (gray).}
  \label{fig:app_scatter_combined}
\end{figure}

Figure~\ref{fig:app_scatter_combined} plots all 115 tasks in lines-changed vs.\ files-changed space.
The leftmost panel shows the full benchmark with dashed reference lines at the medians (3{,}714 lines, 51 files).
The remaining five panels facet the data by programming language, highlighting each language against the full benchmark.
The strong positive correlation confirms that tasks requiring more code also touch more files, and the spread over two orders of magnitude in both dimensions demonstrates the benchmark's diversity.
Python tasks cluster in a moderate range with several high-complexity outliers, while Go and Rust tasks tend toward high file counts due to generated code and macro expansions.

\begin{figure}[h]
  \centering
  \includegraphics[width=\linewidth]{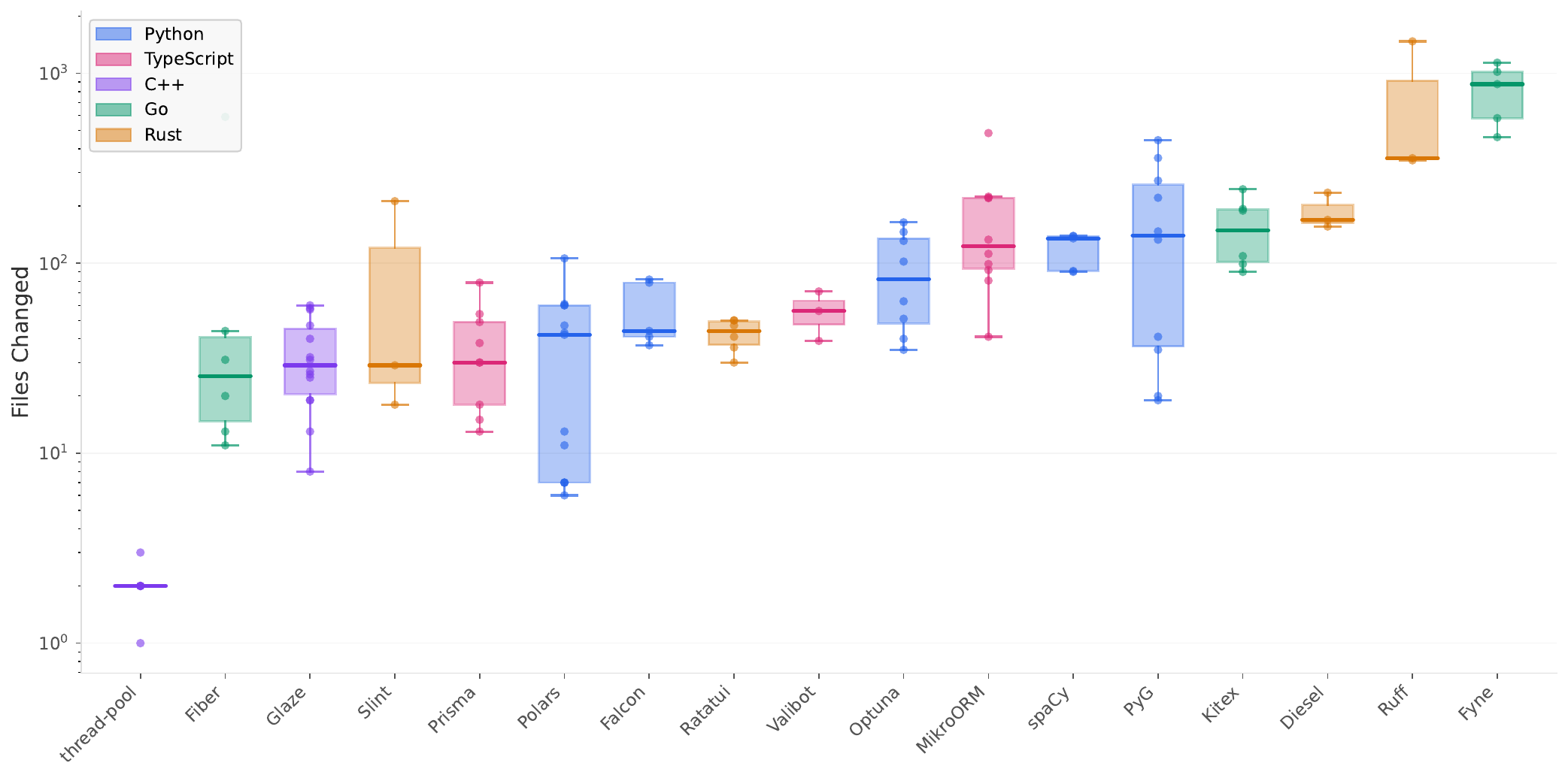}
  \caption{\textbf{Distribution of files changed (oracle patch) across repositories.}
  Repos are sorted by median files changed (log scale); individual task values are shown as jittered points.}
  \label{fig:app_files}
\end{figure}

Figure~\ref{fig:app_files} presents vertical boxplots of files changed per repository, sorted by median.
The log-scale y-axis highlights that complexity varies by more than two orders of magnitude across the benchmark.


\subsection{Temporal Span and Repository Scale}

\begin{figure}[h]
  \centering
  \includegraphics[width=\linewidth]{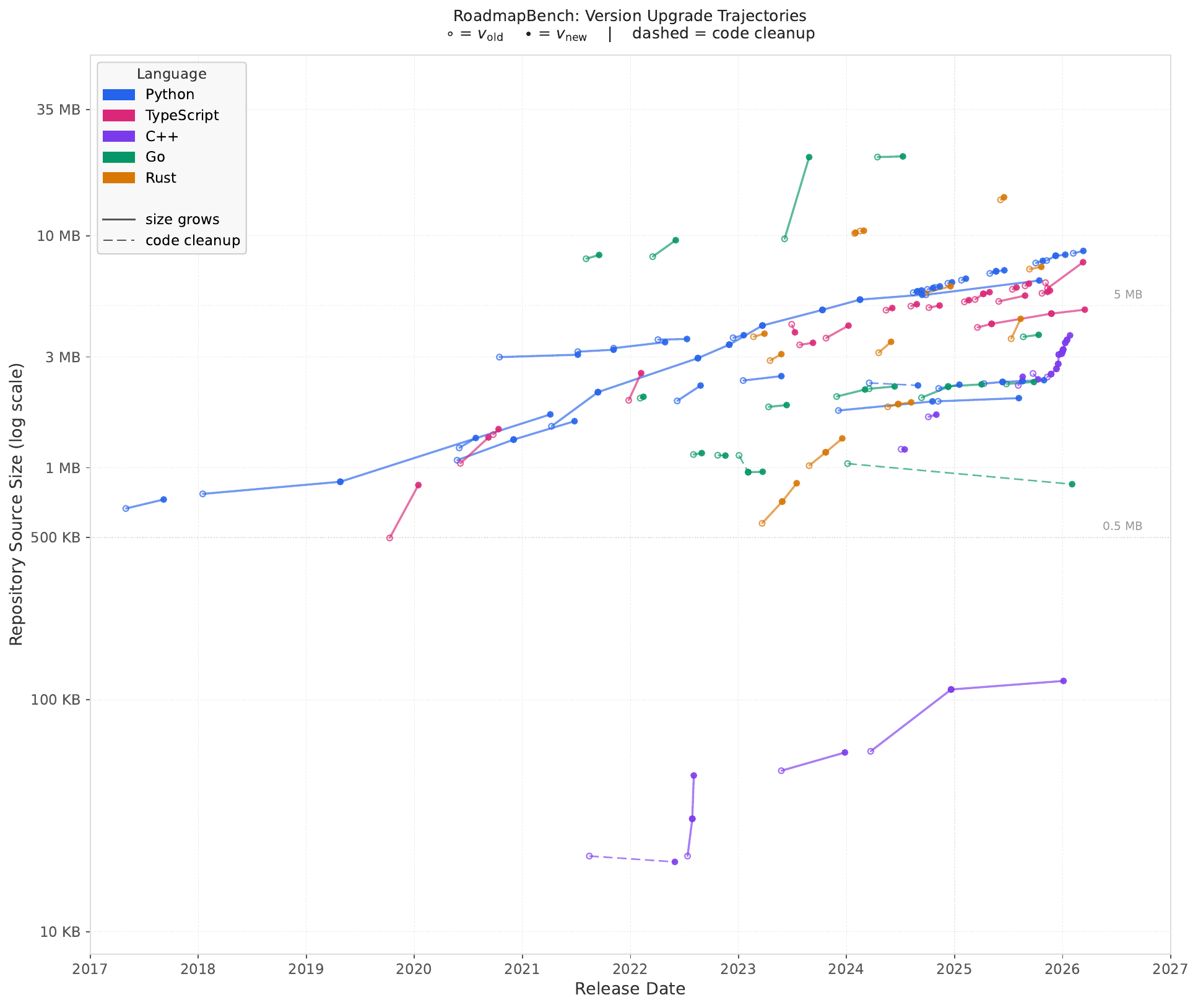}
  \caption{\textbf{Version upgrade trajectories.}
  Each segment represents one task: hollow circles mark the source version, filled dots mark the target version.
  Solid lines indicate codebase growth; dashed lines indicate net size reduction.
  The temporal spread (2017--2026) and size diversity (20\,KB--10\,MB) demonstrate broad benchmark coverage.}
  \label{fig:app_constellation}
\end{figure}

Figure~\ref{fig:app_constellation} visualizes the version upgrade trajectories as a constellation plot.
Each line segment connects a task's source version release to its target version, positioned by release date (x-axis) and repository source size (y-axis, log scale).
Solid lines indicate that the codebase grew between versions; dashed lines indicate code cleanup (net size reduction).

The benchmark spans releases from 2017 to 2026, covering nearly a decade of software evolution.
Repository sizes range from ${\sim}20$\,KB (thread-pool) to ${\sim}10$\,MB (Polars, Go repositories), demonstrating diversity across small libraries and large codebases.
The temporal spread reduces the risk of memorization from training data.

\clearpage
\section{Task Example}
\label{app:examples}

Below is the complete instruction for \texttt{opt-4.0.0-roadmap}, a representative \MYBENCH{} task grounded in the real Optuna \texttt{v3.6.0}$\to$\texttt{v4.0.0} transition (164-day window, oracle patch: 164 files, 7{,}794 LOC filtered).

\newenvironment{targetblk}[2]{%
  \begin{tcolorbox}[phasebox,
    before={\vspace{2pt}},
    top=1pt, bottom=2pt, left=4pt, right=4pt,
    title={\textbf{Target #1}\hfill\normalfont\footnotesize\itshape #2}]%
}{%
  \end{tcolorbox}%
}

\begin{tcolorbox}[
  taskbox,
  label={fig:task-example},
  title={\small\textbf{Hyperparameter Optimization Framework Development Roadmap}},
  top=3pt, bottom=3pt, left=5pt, right=5pt,
  fontupper=\footnotesize,
]

\noindent\textbf{Overview.}\enspace
This library is a hyperparameter optimization framework widely used in machine learning and scientific computing. It provides automated search over parameter spaces using efficient sampling algorithms, distributed optimization via various storage backends, and rich visualization of optimization results.

\smallskip
\noindent\textbf{Goals.}\enspace
Stabilize two experimental subsystems: the artifact management system and the journal-based distributed storage backend. Introduce constrained optimization awareness into \texttt{best\_trial} and \texttt{best\_trials}. Add two new termination components --- \texttt{EMMREvaluator} and \texttt{MedianErrorEvaluator}. Add \texttt{is\_exhausted()} to the grid search sampler.

\vspace{2pt}\tcbline\vspace{1pt}

\begin{targetblk}{1: Artifact Store Official APIs}{}

Objective functions often produce files (model snapshots, logs) that need tracking alongside trial metadata. This target stabilizes the upload API and introduces \texttt{download\_artifact} and \texttt{get\_all\_artifact\_meta}.

\smallskip
\noindent\textbf{Requirements}
\begin{enumerate}[leftmargin=1.4em, itemsep=0pt, topsep=1pt, parsep=0pt, label=\arabic*.]
  \item \texttt{ArtifactMeta} --- frozen dataclass importable from \texttt{optuna.artifacts} with fields: \texttt{artifact\_id: str}, \texttt{filename: str}, \texttt{mimetype: str}, \texttt{encoding: str | None}.

  \item \texttt{download\_artifact} --- importable from \texttt{optuna.artifacts}. All parameters keyword-only: \texttt{artifact\_store} (\texttt{ArtifactStore}), \texttt{file\_path} (\texttt{str}), \texttt{artifact\_id} (\texttt{str}). Returns \texttt{None}. Raises \texttt{FileExistsError} if \texttt{file\_path} already exists.

  \item \texttt{get\_all\_artifact\_meta} --- importable from \texttt{optuna.artifacts}. Positional: \texttt{study\_or\_trial} (\texttt{Trial}, \texttt{FrozenTrial}, or \texttt{Study}). Keyword-only: \texttt{storage} (default \texttt{None}, required for \texttt{FrozenTrial}). Returns \texttt{list[ArtifactMeta]}. Raises \texttt{ValueError} if \texttt{storage} is \texttt{None} and input is a \texttt{FrozenTrial}. When given a \texttt{Study}, returns only study-level artifacts.

  \item \texttt{upload\_artifact} --- updated to use keyword-only parameters in order: \texttt{artifact\_store}, \texttt{file\_path}, \texttt{study\_or\_trial}, with \texttt{storage}, \texttt{mimetype}, \texttt{encoding} as additional keyword-only. Backward compatibility with old positional order maintained. Returns \texttt{str} (artifact ID). Infers MIME type from file extension; defaults to \texttt{"application/octet-stream"}.

  \item \texttt{optuna.artifacts.\_\_all\_\_} must include: \texttt{ArtifactMeta}, \texttt{FileSystemArtifactStore}, \texttt{Boto3ArtifactStore}, \texttt{GCSArtifactStore}, \texttt{Backoff}, \texttt{get\_all\_artifact\_meta}, \texttt{upload\_artifact}, \texttt{download\_artifact}.
\end{enumerate}

\end{targetblk}

\begin{targetblk}{2: JournalStorage API Reorganization}{}

The journal-based storage backend enables distributed optimization over NFS by recording operation logs instead of state snapshots. The module is being reorganized from a private location to a public subpackage with clearer naming conventions.

After this target, users should import journal components from \texttt{optuna.storages.journal} using the new class names, while old names remain available (with deprecation warnings) from \texttt{optuna.storages}.

\smallskip
\noindent\textbf{Requirements}
\begin{enumerate}[leftmargin=1.4em, itemsep=0pt, topsep=1pt, parsep=0pt, label=\arabic*.]
  \item Create public subpackage \texttt{optuna/storages/journal/} containing: \texttt{\_\_init\_\_.py}, \texttt{\_base.py}, \texttt{\_file.py}, \texttt{\_redis.py}, \texttt{\_storage.py}.

  \item Class renames (importable from \texttt{optuna.storages.journal}): \texttt{BaseJournalBackend} (was \texttt{BaseJournalLogStorage}), \texttt{BaseJournalSnapshot} (was \texttt{BaseJournalLogSnapshot}), \texttt{JournalFileBackend} (was \texttt{JournalFileStorage}), \texttt{JournalRedisBackend} (was \texttt{JournalRedisStorage}), \texttt{JournalFileSymlinkLock}, \texttt{JournalFileOpenLock}, \texttt{JournalStorage} (unchanged).

  \item Old names remain importable from \texttt{optuna.storages} with deprecation warnings. \texttt{BaseJournalLogStorage} should subclass \texttt{BaseJournalBackend} decorated with \texttt{@deprecated\_class}.

  \item \texttt{optuna.storages.journal.\_\_all\_\_} must include: \texttt{JournalFileBackend}, \texttt{BaseJournalBackend}, \texttt{JournalFileOpenLock}, \texttt{JournalFileSymlinkLock}, \texttt{JournalRedisBackend}, \texttt{JournalStorage}.
\end{enumerate}

\end{targetblk}

\begin{targetblk}{3: Constrained Optimization in Study Properties}{}

When running constrained optimization, users set constraint values on each trial via system attributes. However, \texttt{best\_trial} and \texttt{best\_trials} currently ignore these constraints. This target makes these properties constraint-aware.

After this target, \texttt{study.best\_trial} returns the best feasible trial (all constraint values $\leq 0.0$), and \texttt{study.best\_trials} computes the Pareto front from only feasible trials. If no feasible trials exist, \texttt{best\_trial} raises \texttt{ValueError}.

\smallskip
\noindent\textbf{Requirements}
\begin{enumerate}[leftmargin=1.4em, itemsep=0pt, topsep=1pt, parsep=0pt, label=\arabic*.]
  \item Helper module \texttt{optuna/study/\_constrained\_optimization.py}: define constant \texttt{\_CONSTRAINTS\_KEY = "constraints"}. Implement \texttt{\_get\_feasible\_trials(trials)} returning only trials where all constraint values are $\leq 0.0$. Trials without a \texttt{"constraints"} key are considered infeasible.

  \item \texttt{best\_trial} property: if the best trial is infeasible, filter to feasible trials and select the one with best objective value (respecting \texttt{study.direction}). Raise \texttt{ValueError} if none exist.

  \item \texttt{best\_trials} property: when any trial has the constraints key, compute Pareto front from feasible trials only.
\end{enumerate}

\end{targetblk}

\begin{targetblk}{4: New Terminator Algorithms}{}

The existing termination framework allows optimization to stop when further trials are unlikely to yield improvements. This target introduces two new components: \texttt{EMMREvaluator} (Expected Minimum Model Regret) and \texttt{MedianErrorEvaluator} (derives threshold from paired improvement evaluator's outputs).

After this target, a user can create an \texttt{EMMREvaluator}, pair it with a \texttt{MedianErrorEvaluator}, and pass both to a \texttt{Terminator} for GP-based automatic stopping.

\smallskip
\noindent\textbf{Requirements}
\begin{enumerate}[leftmargin=1.4em, itemsep=0pt, topsep=1pt, parsep=0pt, label=\arabic*.]
  \item \texttt{EMMREvaluator} --- importable from \texttt{optuna.terminator}, inherits \texttt{BaseImprovementEvaluator}. Constructor: \texttt{\_\_init\_\_(self, deterministic\_objective=False, delta=0.1, min\_n\_trials=2, seed=None)}. Raises \texttt{ValueError} if \texttt{min\_n\_trials <= 1}. Method \texttt{evaluate}: returns EMMR value; returns \texttt{sys.float\_info.max} with insufficient trials or empty search space.

  \item \texttt{MedianErrorEvaluator} --- importable from \texttt{optuna.terminator}, inherits \texttt{BaseErrorEvaluator}. Constructor: \texttt{\_\_init\_\_(self, paired\_improvement\_evaluator, warm\_up\_trials=10, n\_initial\_trials=20, threshold\_ratio=0.01)}. Raises \texttt{ValueError} for invalid args. Method \texttt{evaluate}: before sufficient data returns negative sentinel; on first sufficient call computes median of improvement values multiplied by \texttt{threshold\_ratio}, caches result.

  \item \texttt{optuna.terminator.\_\_all\_\_} must include both \texttt{EMMREvaluator} and \texttt{MedianErrorEvaluator}.
\end{enumerate}

\end{targetblk}

\begin{targetblk}{5: GridSampler Exhaustion Check}{}

When using \texttt{GridSampler}, the user may want to programmatically check whether all parameter combinations have been evaluated. Currently there is no public API for this.

\smallskip
\noindent\textbf{Requirement}
\begin{enumerate}[leftmargin=1.4em, itemsep=0pt, topsep=1pt, parsep=0pt, label=\arabic*.]
  \item \texttt{is\_exhausted(self, study: Study) -> bool} on \texttt{GridSampler}: returns \texttt{True} if all grid combinations have been evaluated, \texttt{False} otherwise.
\end{enumerate}

\end{targetblk}

\begin{tcolorbox}[phasebox, title={\textbf{Completion Criteria}}]
\begin{itemize}[leftmargin=1.2em, itemsep=0pt, topsep=0pt, parsep=0pt]
  \item All new classes and functions importable from their documented paths
  \item Existing APIs remain unchanged (backward compatibility)
  \item Deprecated old names still importable with deprecation warnings
  \item Constraint-aware \texttt{best\_trial} raises \texttt{ValueError} when no feasible trials exist
  \item \texttt{EMMREvaluator} returns finite values with sufficient trials and large values with insufficient data
  \item \texttt{MedianErrorEvaluator} caches its threshold after first computation
  \item \texttt{GridSampler.is\_exhausted()} correctly reports grid coverage
\end{itemize}
\end{tcolorbox}

\end{tcolorbox}
\clearpage
\section{Construction Pipeline Details}
\label{app:pipeline}

\subsection{Repository Selection Criteria}
\label{app:repo_selection}

Candidate repositories must satisfy the following hard constraints: at least 1{,}000 GitHub stars, five or more tagged releases, continued release activity through 2025, and a primary language among our five targets (Python, TypeScript, Go, Rust, Java).

\paragraph{Definition of high-quality release documentation.}
We require that each selected repository maintains release documentation with sufficient information density to support task construction.
Concretely, a release qualifies as high-quality if it satisfies the following criteria:

\begin{itemize}[leftmargin=*,nosep]
  \item Uses natural language to describe what changed in the version, rather than merely listing pull-request numbers or commit hashes.
  \item Explains the background or motivation behind non-trivial changes (e.g., ``to address X limitation'' or ``in response to user feedback on Y'').
  \item Clearly states user-facing impacts such as breaking changes, deprecated APIs, behavioral modifications, or newly introduced features.
  \item Optionally includes code examples, configuration snippets, or migration guides (these are positive signals but not strictly required).
\end{itemize}

\begin{figure}[h]
  \centering
  \setlength{\fboxsep}{0pt}
  \setlength{\fboxrule}{0.3pt}
  \definecolor{imgborder}{gray}{0.65}

  \fcolorbox{imgborder}{white}{\includegraphics[width=0.32\textwidth]{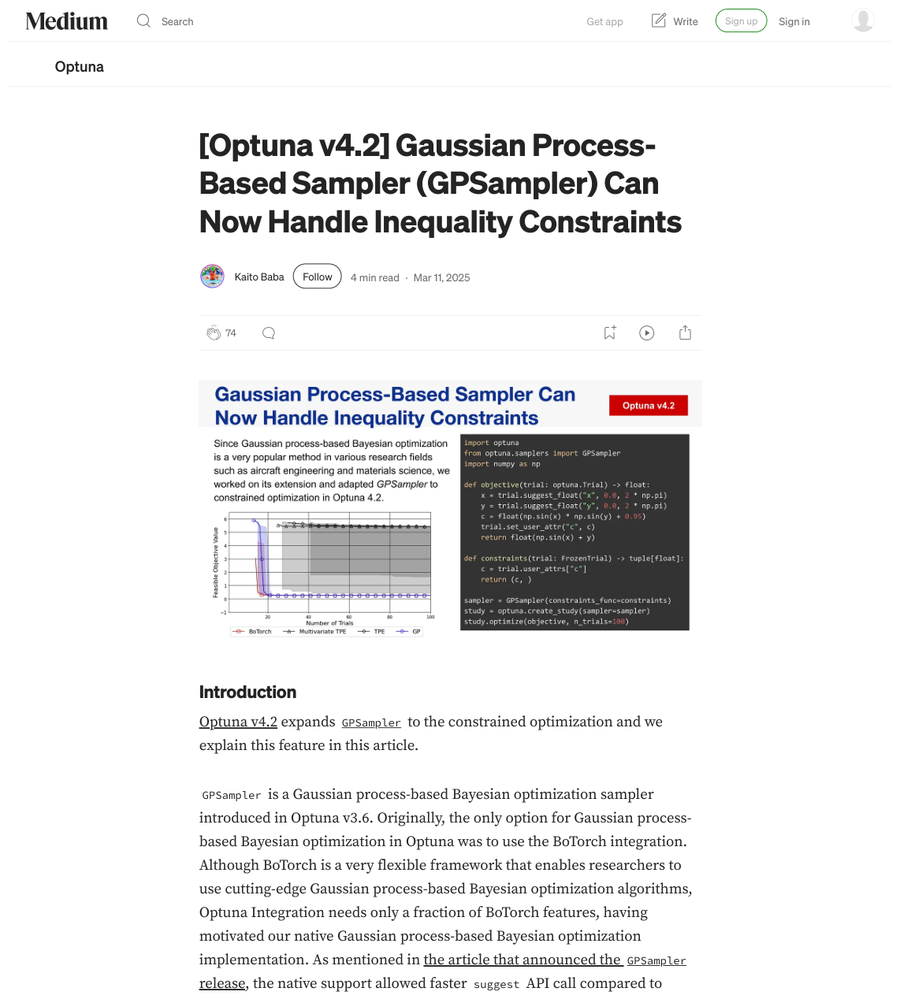}}\hfill
  \fcolorbox{imgborder}{white}{\includegraphics[width=0.32\textwidth]{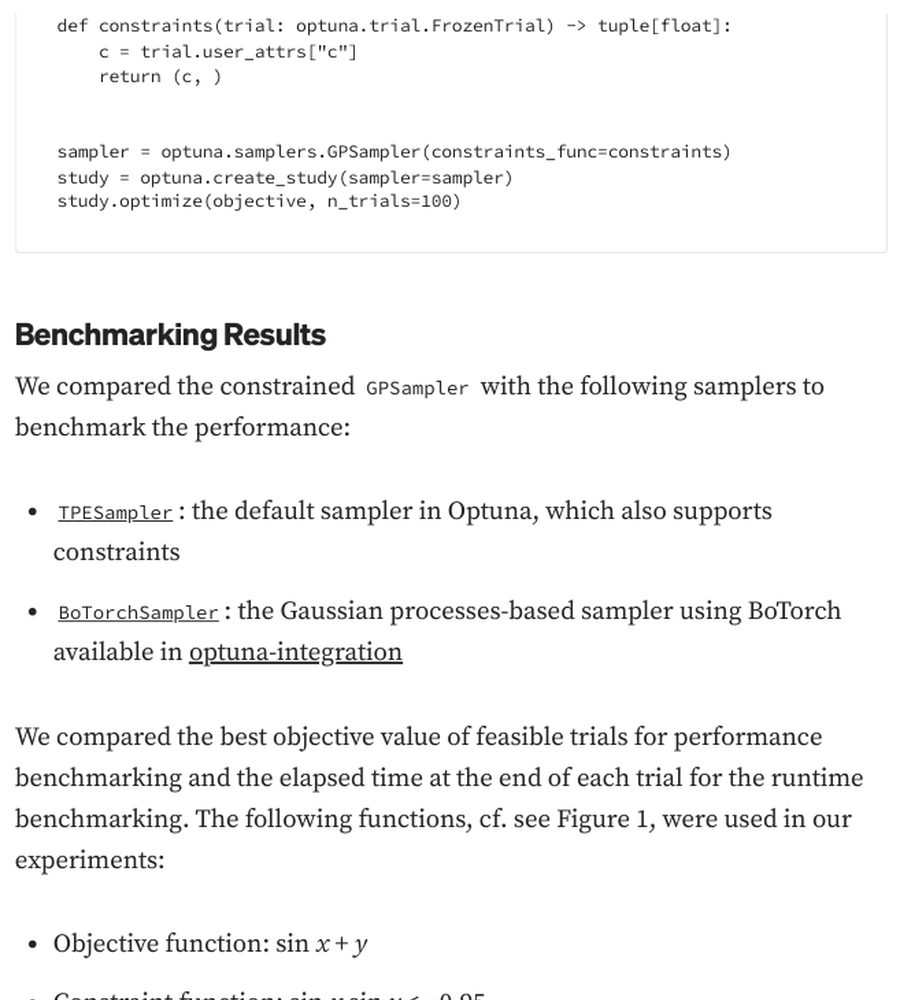}}\hfill
  \fcolorbox{imgborder}{white}{\includegraphics[width=0.32\textwidth]{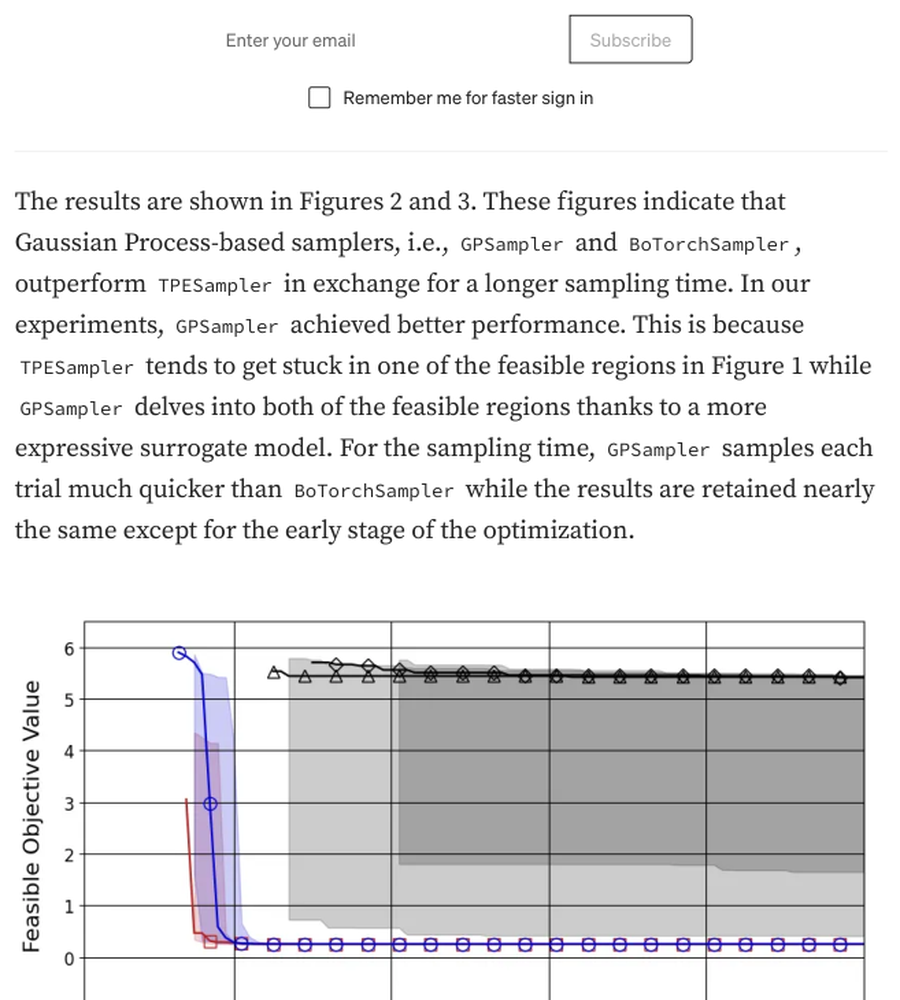}}\\[1pt]
  {\small (a) Optuna v4.2 {\scriptsize(\href{https://medium.com/optuna/optuna-v4-2-gaussian-process-based-sampler-can-now-handle-inequality-constraints-a4f68e8ee810}{medium.com/optuna})}}

  \medskip
  \fcolorbox{imgborder}{white}{\includegraphics[width=0.32\textwidth]{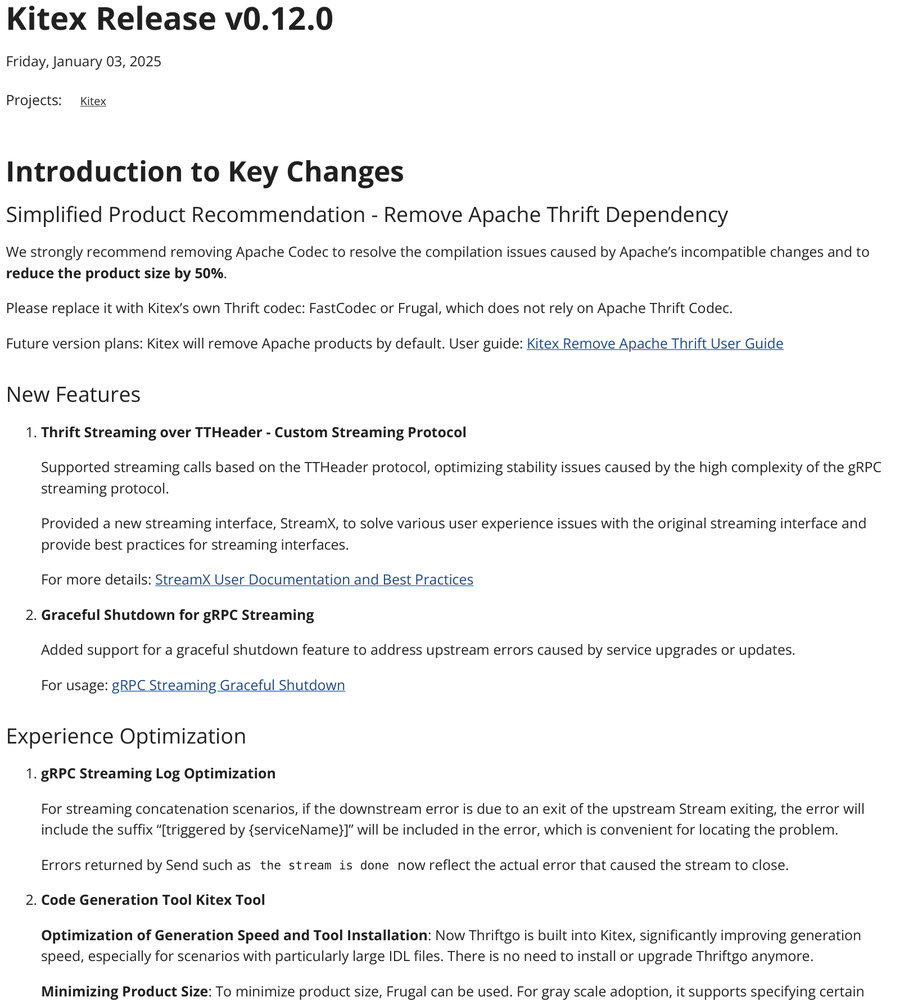}}\hfill
  \fcolorbox{imgborder}{white}{\includegraphics[width=0.32\textwidth]{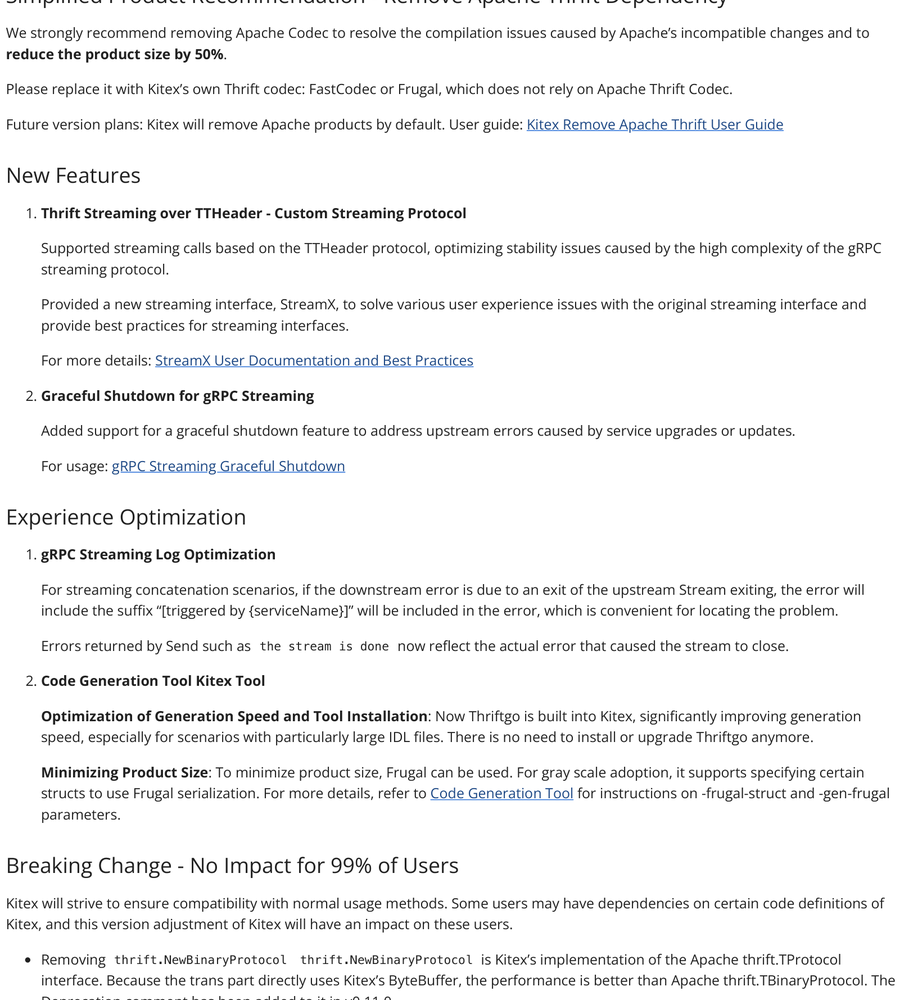}}\hfill
  \fcolorbox{imgborder}{white}{\includegraphics[width=0.32\textwidth]{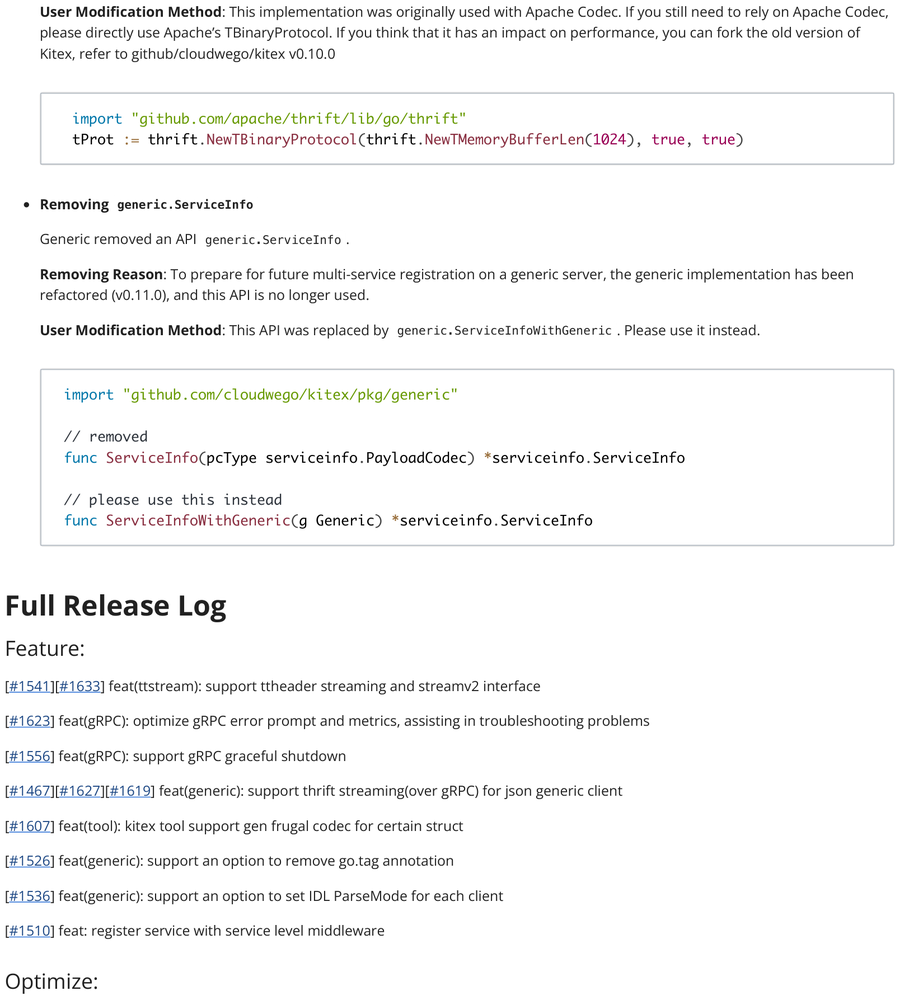}}\\[1pt]
  {\small (b) Kitex v0.12.0 {\scriptsize(\href{https://www.cloudwego.io/zh/blog/2025/01/03/kitex-release-v0.12.0/}{cloudwego.io})}}

  \medskip
  \fcolorbox{imgborder}{white}{\includegraphics[width=0.32\textwidth]{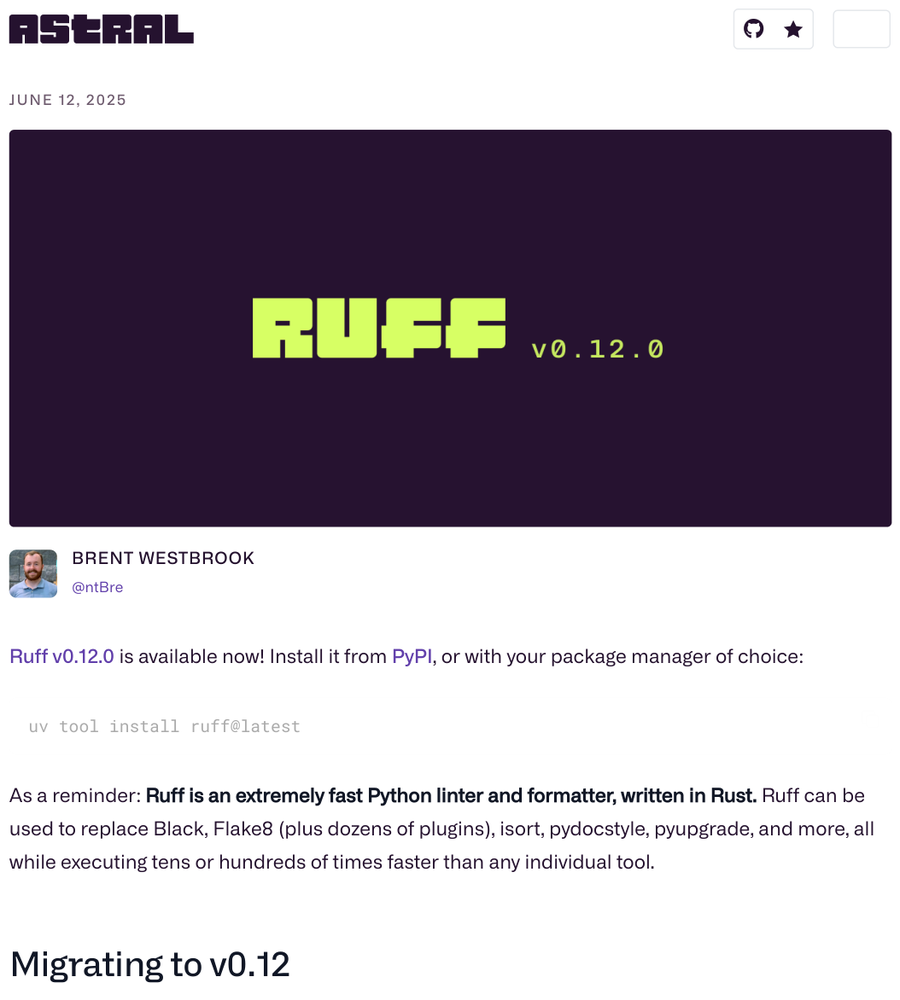}}\hfill
  \fcolorbox{imgborder}{white}{\includegraphics[width=0.32\textwidth]{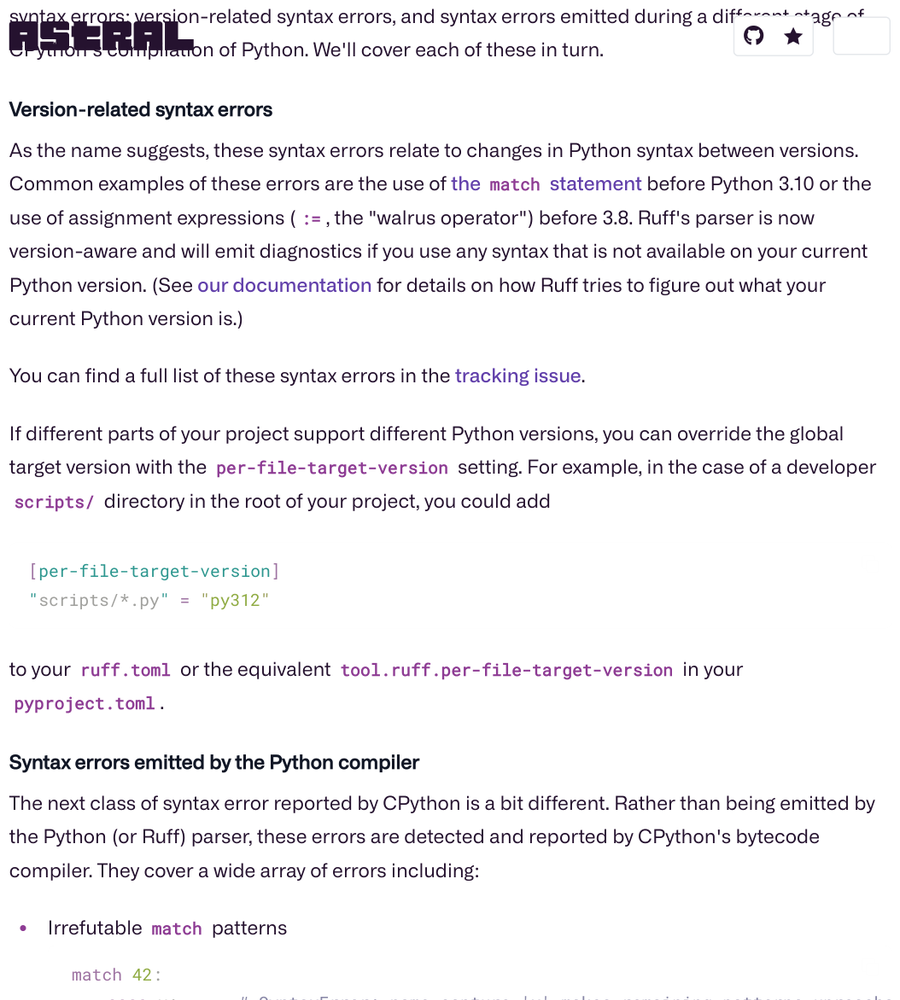}}\hfill
  \fcolorbox{imgborder}{white}{\includegraphics[width=0.32\textwidth]{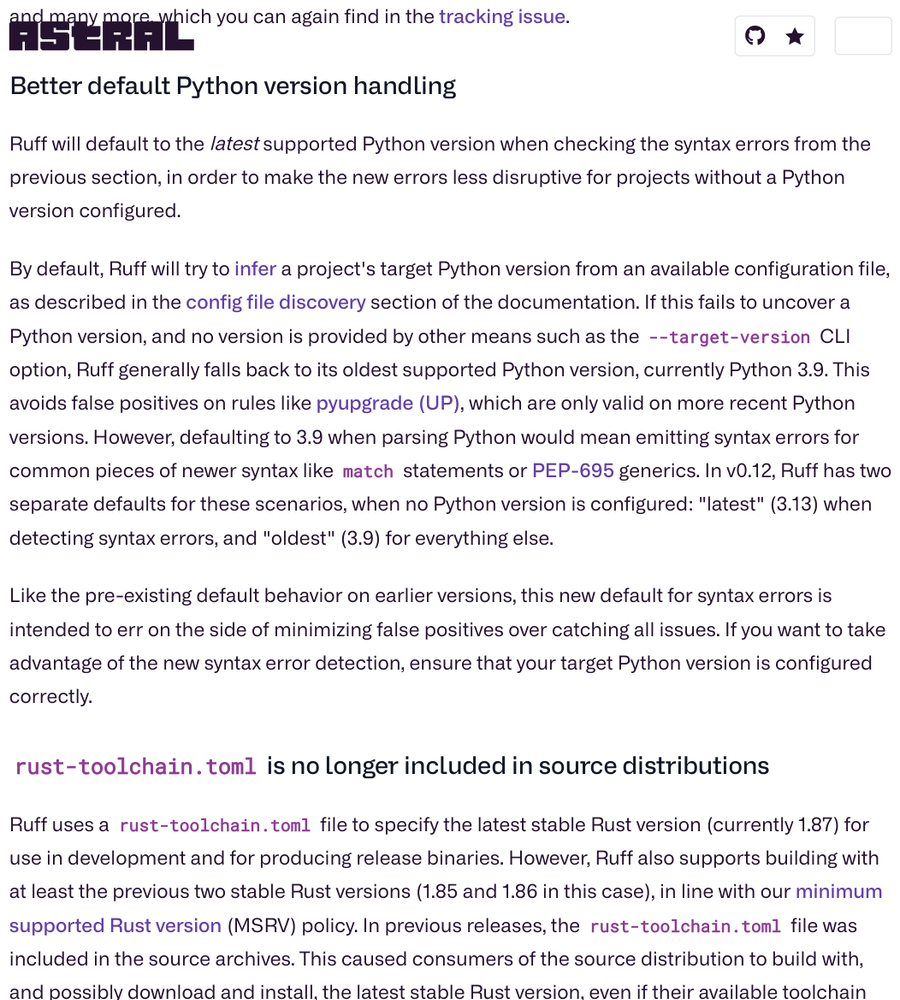}}\\[1pt]
  {\small (c) Ruff v0.12.0 {\scriptsize(\href{https://astral.sh/blog/ruff-v0.12.0}{astral.sh})}}

  \caption{\textbf{Examples of high-quality release documentation from three selected repositories.} Each row shows cropped excerpts from a single version release, illustrating feature narratives, code examples, migration guides, and breaking-change descriptions that serve as source material for task construction.}
  \label{fig:release_examples}
\end{figure}

\noindent Releases that consist solely of auto-generated commit lists (e.g., \texttt{fix \#123}, \texttt{merge PR \#456}), empty bodies, or single-line descriptions are excluded.
Each repository must have at least three releases meeting the above standard.
Figure~\ref{fig:release_examples} shows representative examples of qualifying release documentation.

\paragraph{Expert review and version-pair selection.}
Expert reviewers verify the quality of release documentation identified in the previous step and select consecutive version pairs suitable for task construction.
A version pair is retained if it satisfies: (1) a non-trivial code delta of at least 500 lines changed, (2) at least one externally visible behavioral change expressible as a deterministic test, and (3) release documentation that describes the change in sufficient detail to construct an instruction.

\subsection{Static Review Details}
\label{app:static_review}

Stage 3 applies two complementary reviews under structured checklists, with expert reviewers assisted by Claude-Opus-4.7.

\paragraph{Compliance review (20 items).}
The compliance review is conducted from the perspective of a solver who has \emph{no prior knowledge} of the repository or version upgrade.
It covers five categories:

\begin{enumerate}[leftmargin=*,nosep]
  \item \textbf{Specification clarity} (Q1--Q3): each target's goal and constraints are explicitly stated; the instruction is self-contained without referencing the construction process; requirements are defined positively rather than by exclusion.
  \item \textbf{Implementation leakage} (Q4): a systematic scan for five leakage types---algorithm/flow steps, internal naming, pseudo-code control flow, bug root-cause disclosure, and refactoring checklists---that reveal \emph{how} to implement rather than \emph{what} behavior is required.
  \item \textbf{Information integrity} (Q5--Q7): public API contracts are unambiguous; no test metadata (file names, function names, scoring details) is disclosed; no version numbers or repository names appear.
  \item \textbf{Narrative quality} (Q8--Q9): the instruction provides a coherent version narrative with clear priority ordering among targets; individual target sections follow a consistent structure (background, requirements, constraints).
  \item \textbf{Test conventions} (T1--T7): tests use the required directory layout; target weights sum to 1.0; tests are deterministic and environment-independent; tests do not check implementation internals beyond the specified public contract.
\end{enumerate}

\noindent A task fails the compliance review if any item is marked \textsc{fail}.
The synthesis agent revises the instruction or tests accordingly and re-validates.

\paragraph{Per-target correctness review.}
For each target independently, a reviewer checks instruction--test alignment along four dimensions:

\begin{enumerate}[leftmargin=*,nosep]
  \item \textbf{Completeness}: every behavior asserted by tests is stated in the instruction.
  \item \textbf{Faithfulness}: tests do not assert behaviors beyond the instruction specification.
  \item \textbf{Fairness}: tests do not rely on unstated assumptions (e.g., exact error wording, internal names).
  \item \textbf{Minimality}: tests performing only dead-letter matching without behavioral value are flagged for removal.
\end{enumerate}

\noindent Issues are classified as \emph{T-missing}, \emph{T-ambiguous}, \emph{T-incorrect}, or \emph{T-other}.
Each confirmed issue is repaired by updating the instruction or test, and the oracle patch is re-run to confirm the fail-to-pass guarantee.

\subsection{Quality Control Protocol}
\label{app:qc}

\subsubsection{Attribution Classification}

During rollout-based quality control, agent failures are attributed to either task-side defects (T-type) or model-side failures (M-type). T-type defects indicate problems in the task itself, such as missing specifications or flawed tests, while M-type failures reflect genuine limitations of the agent. Attribution is performed through expert review of agent trajectories and test outcomes.

T-type defects are classified into four subcategories: (1) instruction gaps (a behavioral requirement is not mentioned in the instruction), (2) test brittleness (a test assertion is stricter than the instruction warrants, e.g., checking internal implementation details), (3) environment issues (a dependency or environment variable required for the task is missing from the Docker image), and (4) grading errors (the subtask-level test runner assigns incorrect weights or groupings).

M-type failures are classified into three subcategories: (1) design failures (the agent's implementation does not match the specification at a structural level), (2) implementation bugs (the implementation is structurally correct but contains code errors), and (3) debugging failures (the agent identifies an error but fails to correct it within the turn budget).



\subsubsection{Inter-Annotator Agreement}

To assess attribution consistency, we randomly sampled 40 agent trajectories for independent annotation by two annotators. Cohen’s $\kappa$ for T-type vs.\ M-type classification was 0.83, indicating strong agreement. Disagreements were resolved by a third annotator. The classification rubric and calibration examples are included in the supplementary materials.

\subsubsection{Iterative QC Impact}

Each task undergoes iterative validation: an initial rollout identifies T-type defects, which are then fixed before re-evaluation.
Of the 115 tasks, 45 required at least one fix round (average 3.1 rounds).
Table~\ref{tab:qc_impact} reports model performance before and after QC on all 115 tasks.
For tasks that required no fix, the before and after scores are identical.

\begin{table}[H]
  \caption{Impact of iterative QC on model performance (Terminus, 115 tasks). ``Before'': initial validation; ``After'': post-repair rollout.}
  \vspace{4pt}
  \label{tab:qc_impact}
  \centering
  \small
  \setlength{\tabcolsep}{5pt}
  \begin{tabular}{@{}lcccccc@{}}
    \toprule
    & \multicolumn{3}{c}{\textbf{Completion Score}} & \multicolumn{3}{c}{\textbf{Resolved (\%)}} \\
    \cmidrule(lr){2-4} \cmidrule(lr){5-7}
    \textbf{Model} & \textbf{Before} & \textbf{After} & \textbf{$\Delta$} & \textbf{Before} & \textbf{After} & \textbf{$\Delta$} \\
    \midrule
    Claude-Opus-4.6 & 0.564 & 0.683 & \textcolor{green!50!black}{+0.118} & 19.1 & 30.4 & \textcolor{green!50!black}{+11.2} \\
    GLM-5.1         & 0.475 & 0.511 & \textcolor{green!50!black}{+0.036} & 17.4 & 20.4 & \textcolor{green!50!black}{+3.0} \\
    Kimi-K2.5       & 0.329 & 0.348 & \textcolor{green!50!black}{+0.019} &  6.1 &  7.1 & \textcolor{green!50!black}{+1.0} \\
    \bottomrule
  \end{tabular}
\end{table}

\clearpage
\section{Error Classification Details}
\label{app:error_cases}

This appendix provides the complete error taxonomy, classification methodology, per-model distributions, and representative case studies referenced in \S\ref{sec:failure_mode}.

\subsection{Classification Methodology}
\label{app:error_method}

Each failed subtask is classified by a Claude-Sonnet-4.6 instance operating in agentic mode via Claude Code.
For each task containing failed subtasks, the classifier:

\begin{enumerate}[leftmargin=*,itemsep=2pt]
\item Reads the complete test output (\texttt{test-stdout.txt}) containing all subtask results.
\item Reads the task specification (\texttt{instruction.md}) to understand requirements.
\item Optionally inspects the agent's final code or greps the trajectory for relevant context.
\item Outputs a structured classification for each failed subtask: category, sub-type, root-cause phrase (English, 2--5 words), and rationale (1--3 sentences with technical detail).
\end{enumerate}

The task-level approach (one classifier call per task, classifying all failed subtasks together) enables cross-subtask awareness---e.g., recognizing that multiple subtasks fail due to the same root compilation error (classified as one primary \emph{Syntax Error} plus cascading failures).

\paragraph{Validation.}
We manually validated 50 randomly sampled classifications across all models and categories.
The automated classifier achieved 88\% exact-match agreement with expert labels at the category level (following the validation protocol of~\cite{jimenez2024swebench}).
Disagreements primarily involved the boundary between \emph{Code Defect} and \emph{Wiring Error}---both are implementation-level failures, so category-level accuracy is higher than sub-type accuracy.

\paragraph{Coverage and cost.}
Classification covers 3{,}603 failed subtasks across 13 models (1{,}065 task groups).
Total cost is approximately \$350.

\subsection{Error Taxonomy}
\label{app:error_taxonomy}

Table~\ref{tab:error_taxonomy_full} defines the five error categories and fourteen sub-types.
Categories are ordered by \emph{failure stage}---from early catastrophic failures (code does not compile) to late subtle failures (code compiles and runs but produces incorrect results).
Within each category, sub-types capture the specific mechanism of failure.

\begin{table*}[h]
\centering
\small
\caption{\textbf{Error taxonomy with category and sub-type definitions.} Categories are ordered by failure stage. ``Freq.'' shows the distribution across all 3{,}603 classified failures.}
\label{tab:error_taxonomy_full}
\begin{tabular}{@{}llp{7.2cm}r@{}}
\toprule
\textbf{Category} & \textbf{Sub-type} & \textbf{Definition} & \textbf{Freq.} \\
\midrule
\multirow{3}{*}{\shortstack[l]{Build Error\\(28.3\%)}}
  & Cascading & A root error in a shared module causes compilation failure across multiple subtasks. & 15.8\% \\
  & Syntax Error & Direct compilation/linking failure: syntax error, type mismatch, or unresolved symbol. & 11.1\% \\
  & Dependency & Incompatible dependency version or import of an unavailable package. & 1.4\% \\
\midrule
\multirow{2}{*}{\shortstack[l]{Missing Impl.\\(22.6\%)}}
  & Not Implemented & Required functionality entirely absent---symbol or module does not exist. & 13.8\% \\
  & Partially Impl. & Main feature exists but specific sub-requirements are skipped. & 8.8\% \\
\midrule
\multirow{2}{*}{\shortstack[l]{Interface\\Mismatch (6.5\%)}}
  & Wrong Signature & API exists but signature (parameters, return type) does not match. & 3.7\% \\
  & Wrong Path & Code exists but is inaccessible: wrong module path or missing re-export. & 2.7\% \\
\midrule
\multirow{5}{*}{\shortstack[l]{Impl.\ Error\\(38.5\%)}}
  & Code Defect & Logical bug: wrong formula, off-by-one, nil dereference, incorrect condition. & 23.1\% \\
  & Wiring Error & Components correct but integration broken: params not forwarded, features not activated. & 6.8\% \\
  & Misunderstanding & Agent misinterprets the specification; implements wrong semantics. & 5.9\% \\
  & Edge Case & Main path works; failures only on unusual boundary inputs. & 1.4\% \\
  & Runtime Crash & Compiles but crashes at runtime: unhandled exception, deadlock, OOM. & 1.4\% \\
\midrule
\multirow{2}{*}{\shortstack[l]{Agent Failure\\(4.0\%)}}
  & Abandoned & Agent stops working: gives up, analysis paralysis, or skips remaining subtasks. & 3.5\% \\
  & Exhausted & Budget/step/time limit hit or OOM-killed before completion. & 0.5\% \\
\bottomrule
\end{tabular}
\end{table*}

\subsection{Per-Model Error Distribution}
\label{app:error_overall}

\begin{figure}[h]
  \centering
  \includegraphics[width=0.65\linewidth]{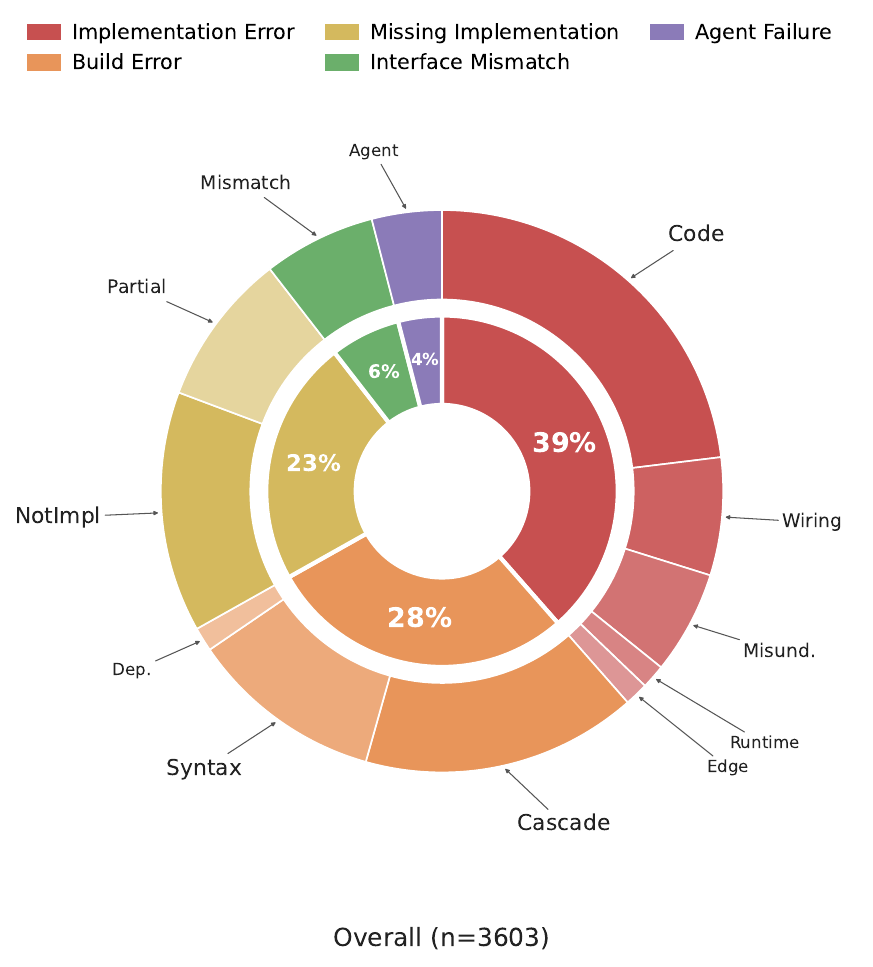}
  \caption{\textbf{Overall error distribution across all models} (n=3{,}603 failed subtasks). Implementation Error dominates (39\%), with Code Defect as the single largest sub-type.}
  \label{fig:error_donut_overall}
\end{figure}

Figure~\ref{fig:error_donut_overall} shows the aggregate error distribution across all 3{,}603 classified failures.
Implementation Error is the dominant category (39\%), followed by Build Error (28\%) and Missing Implementation (23\%).
Within Implementation Error, Code Defect alone accounts for over half of the sub-type (23\% overall).

\begin{figure*}[h]
  \centering
  \includegraphics[width=\linewidth]{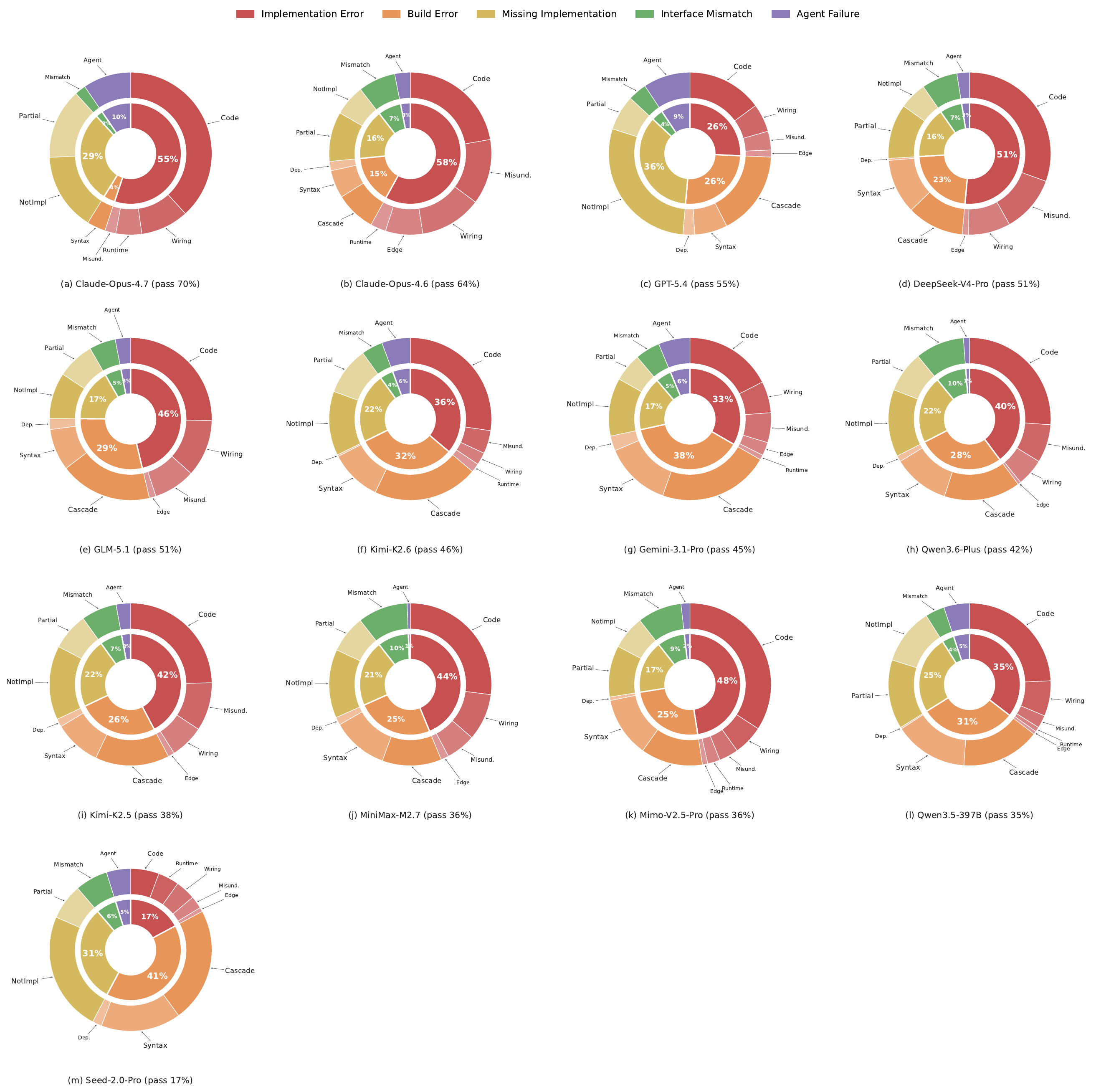}
  \caption{\textbf{Error distribution for all thirteen analyzed models} (inner ring: category proportions; outer ring: sub-type breakdown). Models are ordered by decreasing subtask pass rate from (a) to (m). The dominant failure mode shifts from Implementation Error (strong models) to Build Error and Missing Implementation (weak models).}
  \label{fig:error_donut_all_models}
\end{figure*}

Figure~\ref{fig:error_donut_all_models} breaks this down per model, ordered by subtask pass rate.
Table~\ref{tab:error_per_model} provides the exact counts and percentages.

\begin{table}[h]
\centering
\small
\setlength{\tabcolsep}{7pt}
\renewcommand{\arraystretch}{1.12}
\caption{\textbf{Per-model error category distribution} (count and percentage of failed subtasks). Parentheses after model names indicate subtask pass rate.}
\label{tab:error_per_model}
\begin{tabular}{@{}lcccccr@{}}
\toprule
\textbf{Model} & \textbf{Impl.} & \textbf{Build} & \textbf{Miss.} & \textbf{Intf.} & \textbf{Agent} & \textbf{Total} \\
\midrule
Claude-Opus-4.7 (70\%)   & 75 (55\%) & 5 (4\%) & 40 (29\%) & 3 (2\%) & 13 (10\%) & 136 \\
Claude-Opus-4.6 (64\%)   & 94 (58\%) & 25 (15\%) & 26 (16\%) & 12 (7\%) & 5 (3\%) & 162 \\
GPT-5.4 (55\%)    & 55 (26\%) & 55 (26\%) & 76 (36\%) & 8 (4\%)  & 20 (9\%) & 214 \\
DeepSeek-V4-Pro (51\%)   & 123 (51\%) & 54 (23\%) & 39 (16\%) & 17 (7\%) & 6 (3\%) & 239 \\
GLM-5.1 (51\%)    & 106 (46\%) & 66 (29\%) & 38 (17\%) & 12 (5\%) & 7 (3\%) & 229 \\
Kimi-K2.6 (46\%)  & 102 (36\%) & 89 (32\%) & 63 (22\%) & 12 (4\%) & 16 (6\%) & 282 \\
Gemini-3.1-Pro (45\%)     & 101 (33\%) & 116 (38\%) & 52 (17\%) & 15 (5\%) & 19 (6\%) & 303 \\
Qwen3.6-Plus (42\%)  & 131 (40\%) & 91 (28\%) & 71 (22\%) & 32 (10\%) & 4 (1\%) & 329 \\
Kimi-K2.5 (38\%)  & 132 (42\%) & 80 (26\%) & 69 (22\%) & 22 (7\%) & 9 (3\%) & 312 \\
MiniMax-M2.7 (36\%)    & 141 (44\%) & 79 (25\%) & 68 (21\%) & 32 (10\%) & 2 (1\%) & 322 \\
Mimo-V2.5-Pro (36\%)       & 135 (48\%) & 71 (25\%) & 48 (17\%) & 25 (9\%) & 5 (2\%) & 284 \\
Qwen3.5-397B (35\%)   & 111 (35\%) & 96 (31\%) & 78 (25\%) & 12 (4\%) & 16 (5\%) & 313 \\
Seed-2.0-Pro (17\%)  & 82 (17\%) & 194 (41\%) & 148 (31\%) & 31 (6\%) & 23 (5\%) & 478 \\
\midrule
\textbf{Overall}  & 1{,}388 (39\%) & 1{,}021 (28\%) & 816 (23\%) & 233 (6\%) & 145 (4\%) & 3{,}603 \\
\bottomrule
\end{tabular}
\end{table}

\subsection{Per-Model Analysis}
\label{app:error_per_model}

\paragraph{Claude-Opus-4.7 (pass 70\%).}
The strongest model with fewest total failures (136). Concentrates 55\% in Implementation Error (Code Defect 38\%), with Build Error nearly absent (4\%). Missing Implementation accounts for 29\%, driven equally by Not Implemented and Partially Implemented.

\paragraph{Claude-Opus-4.6 (pass 64\%).}
Concentrates 58\% of failures in Implementation Error, dominated by Code Defect (38\%).
Build Error is rare (15\%), and nearly half of those are cascading failures from a single root cause.
This model rarely leaves features unimplemented; its bottleneck is execution precision.

\paragraph{GPT-5.4 (pass 55\%).}
Uniquely dominated by Missing Implementation (36\%)---the highest among all models.
Agent Failure is also elevated (9\%, all Abandoned), reflecting the ``analysis paralysis'' pattern where the model explores extensively but never starts writing code.
When it does implement, Build and Implementation Errors are balanced (26\% each).

\paragraph{DeepSeek-V4-Pro (pass 51\%).}
Profile resembles Opus but with more Build Errors (23\% vs.\ 15\%).
Implementation Error remains dominant (51\%), indicating strong architectural planning but less precise execution.
Agent Failure is minimal (3\%).

\paragraph{GLM-5.1 (pass 51\%).}
Similar to DeepSeek with 46\% Implementation Error and 29\% Build Error.
The higher Build Error ratio compared to Opus suggests less robust handling of complex type systems and module structures.

\paragraph{Kimi-K2.6 (pass 46\%).}
Balanced between Implementation Error (36\%) and Build Error (32\%), with cascading failures accounting for 21\% of total.
Agent Failure is moderately elevated (6\%), split between Abandoned (12) and Exhausted (4).
Profile sits between GLM-5.1 and Gemini---stronger than its predecessor K2.5 on Implementation Error but with similar Build Error rates.

\paragraph{Gemini-3.1-Pro (pass 45\%).}
Build Error dominates (38\%)---the highest share among mid-tier models.
Cascading failures are frequent (22\% of total failures), indicating that compilation errors in early subtasks propagate to later subtasks.
Implementation Error is relatively lower (33\%).

\paragraph{Qwen3.6-Plus (pass 42\%).}
Interface Mismatch is notably high (10\%), suggesting difficulty with API surface compliance (export paths, naming conventions).
Otherwise balanced between Implementation Error (40\%) and Build Error (28\%).

\paragraph{Kimi-K2.5 (pass 38\%).}
Distribution closely matches Qwen3.6-Plus.
Missing Implementation (22\%) indicates that this model occasionally abandons complex sub-requirements.

\paragraph{MiniMax-M2.7 (pass 36\%).}
Highest Implementation Error percentage among mid-tier models (44\%), with Interface Mismatch also elevated (10\%).
Agent Failure is nearly zero (1\%), meaning the model always attempts implementation---but frequently produces incorrect results.

\paragraph{Mimo-V2.5-Pro (pass 36\%).}
Implementation Error dominates (48\%), with Code Defect at 34\%---the highest raw Code Defect rate among all models.
Interface Mismatch is elevated (9\%), split between Wrong Signature (15) and Wrong Path (10).
Partially Implemented (29) exceeds Not Implemented (19), indicating the model attempts most features but often delivers incomplete solutions.

\paragraph{Qwen3.5-397B (pass 35\%).}
Balanced across Implementation Error (35\%), Build Error (31\%), and Missing Implementation (25\%).
Syntax Error is notably high within Build Error (46 of 96), suggesting frequent compilation-level mistakes rather than cascading propagation.
Partially Implemented (43) strongly dominates Not Implemented (35), a pattern distinct from weaker models where Not Implemented typically leads.

\paragraph{Seed-2.0-Pro (pass 17\%).}
Dominated by Build Error (41\%) and Missing Implementation (31\%).
Implementation Error accounts for only 17\%---not because the model is precise, but because code often fails to compile before behavioral correctness can be evaluated.
This model represents the weakest capability tier where fundamental code generation is the bottleneck.

\subsection{Representative Case Studies}
\label{app:case_studies}

We present one representative case per error category, selected to demonstrate how each failure type manifests in practice.
Each case includes the target requirement, the key test output, and root-cause analysis.

\subsubsection*{Case 1: Implementation Error (Code Defect)}

\begin{tcolorbox}[breakable, colback=red!3, colframe=red!40, title={\small pyg-1.7.2-roadmap Target 6: Regularization Functions $\mid$ Kimi-K2.5}, fonttitle=\small\bfseries]
\small

\textbf{Target Requirement:}
Implement a \texttt{gini(w: Tensor)} function that computes the Gini coefficient of a 2D weight matrix (row-wise inequality of absolute values, averaged across rows).
A fully sparse row gives Gini close to 1.0; uniform gives 0.0.
For a matrix with row 1 = \texttt{[0,0,0,0]} and row 2 = \texttt{[0,0,0,1000]}, the expected Gini is \texttt{0.5}.

\vspace{4pt}
\textbf{Test Output} (1 failed / 7 total):
\begin{verbatim}
FAILED test_gini_regularization
  assert torch.isclose(result, torch.tensor(0.5))
  AssertionError:
    tensor(0.1250) != tensor(0.5000)
\end{verbatim}

\vspace{2pt}
\textbf{Root Cause:}
The normalization formula is inverted.
The agent wrote \texttt{gini = gini / (n - 1)} instead of the correct \texttt{gini = gini * n / (n - 1)}.
For the test input \texttt{[0,0,0,1000]}: raw Gini = 0.75, correct normalized = $0.75 \times \frac{4}{3} = 1.0$, but the implementation computes $0.75 \div 3 = 0.25$.
Averaging with the all-zero row (Gini = 0) yields $0.125$ instead of the expected $0.5$.

\vspace{2pt}
\textbf{Insight:} Six of seven tests pass---the function exists, compiles, and handles most cases correctly.
The failure is a single arithmetic operator error (\texttt{/} vs.\ \texttt{*}) in a normalization formula, exemplifying the ``execution precision'' bottleneck: models understand the algorithm but make subtle mistakes when translating mathematical specifications to code.
\end{tcolorbox}

\subsubsection*{Case 2: Build Error (Circular Import)}

\begin{tcolorbox}[breakable, colback=orange!5, colframe=orange!50, title={\small fal-1.3.0-roadmap Target 1: Media Framework $\mid$ GPT-5.4}, fonttitle=\small\bfseries]
\small

\textbf{Target Requirement:}
Create a pluggable media handling system: \texttt{BaseHandler} abstract class, \texttt{Handlers} registry mapping content types to handler instances, \texttt{JSONHandler}/\texttt{MessagePackHandler} implementations, and a \texttt{validate(schema)} decorator for JSON Schema validation.
Add \texttt{media} properties on Request/Response for automatic serialization.

\vspace{4pt}
\textbf{Test Output} (0 collected, import error):
\begin{verbatim}
ERROR collecting test_01_media.py
  ImportError while importing test module:
  falcon/__init__.py:32
    -> falcon/api.py:21
    -> falcon/routing/__init__.py:22
    -> falcon/routing/compiled.py:21:
       import falcon.routing.converters
  AttributeError: module 'falcon' has no attribute 'routing'
\end{verbatim}

\vspace{2pt}
\textbf{Root Cause:}
Agent used absolute import \texttt{import falcon.routing.converters} in \texttt{compiled.py}.
This statement requires Python to resolve \texttt{falcon.routing} as an attribute of the \texttt{falcon} module object.
However, at this point in the initialization sequence (\texttt{falcon.\_\_init\_\_} $\to$ \texttt{falcon.api} $\to$ \texttt{falcon.routing.\_\_init\_\_} $\to$ \texttt{compiled.py}), the \texttt{falcon.\_\_init\_\_} module has not finished executing, so the \texttt{routing} attribute has not yet been bound to the \texttt{falcon} module namespace---even though \texttt{falcon/routing/\_\_init\_\_.py} is actively being loaded.
The fix is to use a relative import (\texttt{from . import converters}).
The failure cascades to all 5 subtasks (no tests can be collected).

\vspace{2pt}
\textbf{Insight:} A single import-path mistake renders the entire codebase unimportable, demonstrating how Build Errors---especially cascading ones---produce catastrophic multi-target failures.
\end{tcolorbox}

\subsubsection*{Case 3: Missing Implementation (Not Implemented)}

\begin{tcolorbox}[breakable, colback=yellow!5, colframe=yellow!60!black, title={\small opt-3.2.0-roadmap Target 4: BIPOP CMA-ES $\mid$ Seed-2.0-Pro}, fonttitle=\small\bfseries]
\small

\textbf{Target Requirement:}
Extend \texttt{CmaEsSampler} to accept \texttt{restart\_strategy="bipop"}, implementing BI-population CMA-ES that alternates between large-population and small-population restarts.
Add \texttt{n\_restarts\_with\_large}, \texttt{poptype}, \texttt{small\_n\_eval}, \texttt{large\_n\_eval} fields to the \texttt{\_CmaEsAttrKeys} NamedTuple.
Invalid strategy values must raise \texttt{ValueError}.

\vspace{4pt}
\textbf{Test Output} (16 failed / 17 total):
\begin{verbatim}
FAILED test_bipop_available
  ValueError: restart_strategy=bipop is unsupported.
  Please specify: 'ipop', 'bipop' or None.

FAILED test_sampler_attr_key_bipop[options0-cma:]
  AttributeError: '_CmaEsAttrKeys' object has no
  attribute 'n_restarts_with_large'

FAILED test_restore_optimizer_after_restart_bipop
  ValueError: restart_strategy=bipop is unsupported.
\end{verbatim}

\vspace{2pt}
\textbf{Root Cause:}
Agent left only \texttt{\# TODO(c-bata): Support BIPOP-CMA-ES.} without implementing any functionality.
The \texttt{restart\_strategy} validator still only accepts \texttt{'ipop'} and \texttt{None}; the \texttt{\_CmaEsAttrKeys} NamedTuple was never extended.
The sole passing test (\texttt{test\_invalid\_restart\_strategy}) checks that truly invalid values (e.g., \texttt{'foo'}) raise exceptions---it passes because \texttt{'bipop'} is now also rejected, though it should have been a valid option.

\vspace{2pt}
\textbf{Insight:} Weak models skip complex algorithmic requirements entirely rather than attempting partial implementations, resulting in a pattern of TODO comments as placeholders.
\end{tcolorbox}

\subsubsection*{Case 4: Interface Mismatch (Wrong Export Path)}

\begin{tcolorbox}[colback=green!5, colframe=green!50!black, title={\small fal-3.0.0-roadmap Target 1: ASGI Support $\mid$ Qwen3.6-Plus}, fonttitle=\small\bfseries]
\small

\textbf{Target Requirement:}
Create \texttt{falcon.asgi} package with async App, Request, Response, BoundedStream.
Implement testing utilities (\texttt{ASGIConductor}, \texttt{create\_scope()}, \texttt{SimpleTestResourceAsync}) and sync/async bridge functions (\texttt{sync\_to\_async}, \texttt{async\_to\_sync}).
All must be importable from \texttt{falcon.testing} and the top-level \texttt{falcon} namespace.

\vspace{4pt}
\textbf{Test Output} (12 failed / 43 total):
\begin{verbatim}
FAILED test_asgi_conductor_default_headers
  ImportError: cannot import name 'ASGIConductor'
  from 'falcon.testing'

FAILED test_sync_to_async
  falcon/util/sync.py:21:
  RuntimeError: no running event loop

FAILED test_unsupported_http_version[0.9]
  Failed: DID NOT RAISE UnsupportedError
\end{verbatim}

\vspace{2pt}
\textbf{Root Cause:}
Agent implemented \texttt{ASGIConductor}, \texttt{SimpleTestResourceAsync}, and \texttt{create\_scope} in \texttt{testing/asgi\_client.py}, but \texttt{testing/\_\_init\_\_.py} never imports from that file.
These classes exist on disk but are inaccessible via the \texttt{falcon.testing} namespace that tests use.
Additionally, \texttt{async\_to\_sync} calls \texttt{asyncio.get\_event\_loop()} which fails on Python 3.10+ without a running loop.

\vspace{2pt}
\textbf{Insight:} Writing correct code is necessary but not sufficient---the code must also be properly exported at the expected module path. This class of error is especially common in Python packages with explicit \texttt{\_\_init\_\_.py} re-exports.
\end{tcolorbox}

\subsubsection*{Case 5: Agent Failure (Infinite Loop Until Budget Exhaustion)}

\begin{tcolorbox}[breakable, colback=purple!3, colframe=purple!40, title={\small mko-4.0.0-roadmap Target 1--7: ORM Core Decorators $\mid$ Claude-Opus-4.6}, fonttitle=\small\bfseries]
\small

\textbf{Target Requirement:}
Implement seven core ORM decorators (\texttt{@Filter}, \texttt{@Subscriber}, \texttt{@Embeddable}, \texttt{@Formula}, etc.) with full TypeScript decorator semantics, metadata storage, and integration with the entity manager lifecycle.

\vspace{4pt}
\textbf{Agent Trajectory} (249 steps, infinite loop):
\begin{verbatim}
Step 1-40:   monorepo directory restructuring
Step 41-180: repeated attempts to reorganize packages/
Step 181-245: trapped in a loop:
  Step 245: ls packages/core/src/ | wc -l
  Step 246: ls packages/core/src/ | wc -l
  Step 247: ls packages/core/src/ | wc -l
  ...
  Step 249: (budget exhausted, forced termination)
  === zero decorator implementations written ===
\end{verbatim}

\vspace{4pt}
\textbf{Test Output} (all 7 targets fail identically):
\begin{verbatim}
TypeError: core_1.Filter is not a function
TypeError: core_1.Subscriber is not a function
TypeError: core_1.Embeddable is not a function
  (all decorators undefined - never implemented)
\end{verbatim}

\vspace{2pt}
\textbf{Root Cause:}
The agent spent its entire 249-step budget on monorepo directory restructuring (copying files between directories, checking file counts) without ever beginning to implement any decorator.
In the final 60+ steps, the agent entered a degenerate loop, repeatedly executing the same \texttt{ls | wc -l} command with no progress.
The session was forcibly terminated at the step limit.

\vspace{2pt}
\textbf{Insight:} This is the canonical Agent Failure pattern: the agent gets stuck in preparatory work and never reaches the actual implementation.
Unlike Missing Implementation (Case 3) where a feature is consciously skipped, here the agent \emph{intended} to implement but was trapped in an unproductive loop until its budget was exhausted.
\end{tcolorbox}

\end{document}